\documentclass[
reprint,
superscriptaddress,
amsmath,amssymb,
aps,pra
]{revtex4-2}

\usepackage[utf8]{inputenc}
\usepackage{graphicx}
\usepackage{dcolumn}
\usepackage{bm}
\usepackage{dsfont}
\usepackage{hyperref}
\usepackage[dvipsnames]{xcolor}
\usepackage{amsfonts,amsthm}
\usepackage{xfrac}
\usepackage[normalem]{ulem}

\newcommand{\rr}{\bm{r}}
\newcommand{\eqnref}[1]{Eq.~(\ref{#1})}
\newcommand{\Hop}{\hat{H}}
\newcommand{\w}{\omega}
\newcommand{\bop}{\hat{b}}
\newcommand{\bdop}{\bop^\dag}

\usepackage{soul}

\usepackage{braket}
\graphicspath{{figures/}}

\begin{document}

\title{Atomic waveguide QED with atomic dimers}

\author{David Castells-Graells}\email{david.castells@mpq.mpg.de}
\affiliation{Department of Chemistry, Technical University of Munich, Lichtenbergstrasse 4, 85748 Garching, Germany}
\affiliation{Max-Planck-Institut f\"ur Quantenoptik, Hans-Kopfermann-Strasse 1, 85748 Garching, Germany.}
\affiliation{Munich Center for Quantum Science and Technology, Schellingstrasse 4, 80799 M\"unchen, Germany.}
\author{Daniel Malz}
\affiliation{Max-Planck-Institut f\"ur Quantenoptik, Hans-Kopfermann-Strasse 1, 85748 Garching, Germany.}
\affiliation{Munich Center for Quantum Science and Technology, Schellingstrasse 4, 80799 M\"unchen, Germany.}
\author{Cosimo C. Rusconi}
\affiliation{Max-Planck-Institut f\"ur Quantenoptik, Hans-Kopfermann-Strasse 1, 85748 Garching, Germany.}
\affiliation{Munich Center for Quantum Science and Technology, Schellingstrasse 4, 80799 M\"unchen, Germany.}
\author{J. Ignacio Cirac}
\affiliation{Max-Planck-Institut f\"ur Quantenoptik, Hans-Kopfermann-Strasse 1, 85748 Garching, Germany.}
\affiliation{Munich Center for Quantum Science and Technology, Schellingstrasse 4, 80799 M\"unchen, Germany.}

\date{\today}

\begin{abstract}
Quantum emitters coupled to a waveguide is a paradigm of quantum optics, whose essential properties are described by waveguide quantum electrodynamics (QED).
We study the possibility of observing the typical features of the conventional waveguide QED scenario in a system where the role of the waveguide is played by a one-dimensional subwavelength atomic array.
For the role of emitters, we propose to use anti-symmetric states of atomic dimers -- a pair of closely spaced atoms -- as effective two-level systems, which significantly reduces the effect of free-space spontaneous emission.
We solve the dynamics of the system both when the dimer frequency lies inside and when it lies outside the band of modes of the array.
Along with well-known phenomena of collective emission into the guided modes and waveguide mediated long-range dimer--dimer interactions, we uncover significant non-Markovian corrections which arise from both the finiteness of the array and through retardation effects.
\end{abstract}

\maketitle

\section{Introduction}

The interaction between quantum emitters and a structured reservoir leads to a rich phenomenology. It has recently attracted renewed interest~\cite{chang2018colloquium} due to experimental progress in novel experimental platforms such as superconducting waveguide quantum electrodynamics (QED)~\cite{vanLoo2013,Mirhosseini2018,Mirhosseini2019,Sundaresan2019}, cold atoms near nanofibres~\cite{LeKien2005,Vetsch2010} or photonic crystal waveguides~\cite{Goban2014,Goban2015,Hood2016}, and quantum dots~\cite{Sipahigil2016,Palacios2017,Samkharadze2018}.
The dynamics of emitters coupled to such non-conventional reservoirs is expected to exhibit several distinctive features such as collective super and subradiant emission into the reservoir~\cite{Chang_2012,Goban2015}, long-range dipole--dipole interactions mediated by the reservoir~\cite{john1990quantum,John1991,douglas2015quantum,gonzalez2015subwavelength}, and non-Markovian effects~\cite{gonzalez2017markovian,gonzalez2017quantum,sinha2020non}.
The observation of many of these phenomena is, however, challenging even with the unprecedented level of control achieved today in several experiments. These difficulties arise from both imperfections in the fabrication of these devices as well as from the unavoidable absorption of photons in the waveguide (see \cite{chang2018colloquium} and references therein).

Ordered atomic arrays have recently attracted significant interest as a new paradigm for controlling light--matter interaction~\cite{Mewton2007,porras2008collective,antezza2009fano,antezza2009spectrum,Zoubi2010,bettles2015cooperative,Bettles2016,bettles2016enhanced,asenjo2017atom,asenjo2017exponential,perczel2017photonic,Guimond2019,Bekenstein2020}. Subwavelength arrays, whose interatomic spacing lies below the wavelength of the characteristic atomic dipole transition, exhibit strong collective behaviour such as superradiance and subradiance~\cite{Clemens2003,Mewton2007,Mazets2007,porras2008collective,Bettles2016,Guimond2019,Rui2020,albrecht2019subradiant}.
One can draw an analogy between a one-dimensional atomic array and a conventional waveguide, where collective subradiant excitations can be thought of as propagating modes of the waveguide~\cite{Zoubi2010,asenjo2017exponential,Needham2019}.

Recently, Masson and Asenjo-Garcia have proposed the concept of atomic waveguide QED~\cite{masson2020atomic}.
Inspired by waveguide QED, where optical emitters are coupled to a waveguide, they considered the case in which atoms are coupled to a one-dimensional subwavelength atomic array which plays the role of the waveguide. 
The motivation is that an atomic waveguide in free space is a conceptually simple, clean optical medium that allows, in principle, to eliminate any intrinsic internal losses or imperfections which affect conventional waveguides or photonic crystals, and may feature very low disorder.
The observation of characteristic waveguide QED phenomena in an atomic waveguide setup poses, however, several fundamental challenges~\cite{masson2020atomic}.
On the one hand, efficient coupling between external ``impurity'' atoms and the atomic waveguide is hindered by free space decay due to the presence of superradiant (bright) modes of the array. On the other hand, the dynamics of the emitters shows signs of non-Markovian effects, which ultimately spoil some of the interesting phenomena of conventional waveguide QED.
Some of these difficulties may be overcome by reducing the interatomic separation of the waveguide and by placing the impurity atoms extremely close to the array~\cite{masson2020atomic,patti2020controlling}, but it is challenging to achieve the required deep subwavelength regime experimentally.

In this article, we propose a modification of this setup to mitigate these problems.
Specifically, we propose (i) to use atomic dimers -- a pair of closely spaced atoms -- as effective two-level emitters coupled to the atomic waveguide, and (ii) to control the dimers' linewidth with a Raman transition.
We show that a dimer behaves as an effective two-level system formed by its ground state and its anti-symmetric state. The anti-symmetric state features a reduced coupling to free-space modes and, at the same time, an increased coupling to the array's guided modes.
Additionally, we show that by controlling the decay rate of the dimer atoms via a Raman transition, it is possible to recover a Markovian regime for the dynamics of dimers coupled to an atomic waveguide.
We derive simple models for our setup that predict collective emission from the impurity dimers into an array's subradiant mode, and coherent long-range interactions between impurity dimers mediated by the array. We verify both observations numerically for the case of an atomic array with interatomic separation of quarter-wavelength, a regime where the simpler case of single atoms coupled to an atomic waveguide is hampered by free-space decay and non-Markovian effects~\cite{masson2020atomic}.
We also study the effects and different non-Markovian behaviors arising from the finiteness of the chain and the reduced group velocity at the band edge.
Our results show a clear advantage of using dimers over atoms and highlight a promising route towards observing non-Markovian waveguide QED physics, which has received considerable attention recently \cite{calajo2019exciting,sinha2020non,guo2020oscillating,trivedi2020optimal}.

This article is organised as follows. 
The theoretical model, as well as details on the dimer--array coupling and on the Raman transition used to control the dimer's linewidth are described in Sec.~\ref{sec:System}. The physics of dimers coupled to an atomic array is described in Sec.~\ref{sec:inband}  (Sec.~\ref{sec:bandgap}) for the case of dimer's frequency lying inside (outside) the band of guided modes of the array. We discuss possible generalisations of the case presented here and draw our conclusions in Sec.~\ref{sec:discussion}.
We leave additional details on the derivation to the appendices.

\section{System Description}\label{sec:System}

\begin{figure}[h!]
	\includegraphics[width=0.84\linewidth]{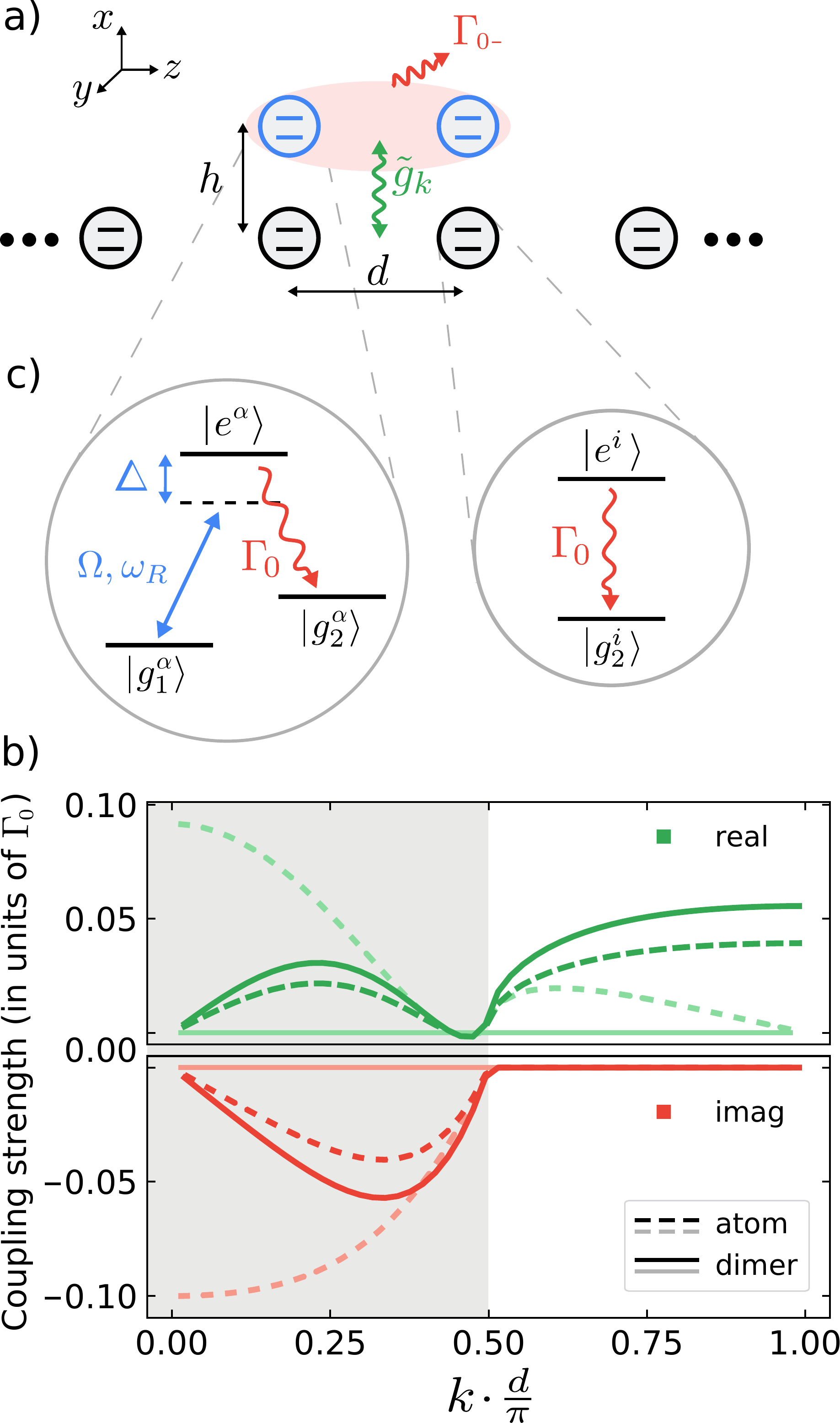}
	\caption{\label{fig1} (a) Schematic representation of the setup studied in the paper. The impurity atoms (blue) interact with collective states of the atomic chain (black).
	(b) Real (top) and imaginary (bottom) part of the coupling of a single impurity atom (dashed) or a dimer in the anti-symmetric, $\lambda=-1$, state (solid), with a chain mode with quasi-momentum $k$. The shaded region indicates the superradiant part of the chain modes. The dimer is aligned with the center, and the single atom with the closest site to the center of an array with $N=100$ and $d=h=\lambda_0/4$. The dimer anti-symmetric state interacts destructively (constructively) with the low(high)-$k$ modes. Hence, in comparison with a single atom, the coupling of the dimer to the guided modes is stronger, while the coupling to the superradiant modes is highly suppressed. The finiteness of the chain adds standing-wave envelopes to the Bloch states, which explains the difference between even and odd modes represented with dark and light lines, respectively. (c) Atomic level scheme of the setup proposed in this work. The two-level impurity atom is realized with a three-level $\Lambda$ atom, in which the transition $\ket{g_1}\to\ket{e}$ is driven by a laser field with detuning $\Delta$. The detuning and spontaneous emission rate are much larger than any other energy in the system and, thus, $\ket{e^\alpha}$ can be eliminated resulting in an effective two-level atom $\ket{g_1^\alpha}$--$\ket{g_2^\alpha}$ (see text for details). The decay rate from $\ket{e^\alpha}$ to $\ket{g_1^\alpha}$ is assumed to be much slower than $\Gamma_0$. The chain modes are diagonalized in $k$-space, with energy $J_k$ and free-space decay $\Gamma_k$.}
\end{figure}

We consider a one-dimensional atomic array of $N$ atoms with resonance frequency $\omega_0$ and lattice spacing $d$, and $n$ impurity atoms with resonance frequency $\omega_0^{\rm imp}$ placed at a distance $h$ from the chain [see Fig.~\ref{fig1}(a)]. We assume all atoms to be polarised along the $z$-axis, which coincides with the direction of the atomic array. The effective dynamics of an atomic ensemble coupled to a continuum of quantized electromagnetic modes was first derived in Ref.~\cite{lehmberg1970radiation}. In the electric-dipole and rotating-wave approximation, and in the Markovian regime, the photonic environment can be eliminated yielding an effective non-Hermitian Hamiltonian ($\hbar=1$)
\begin{equation}\label{eq:fullHam}
	\frac{\hat{H}}{\Gamma_0} = \sum_{i}\frac{\omega_0^i}{\Gamma_0}\,\hat{\sigma}_{ee}^i - 3\pi\sum_{i,j} G_0^{zz}(\bm{r}_i,\bm{r}_j)\,\hat{\sigma}_{eg}^i\hat{\sigma}_{ge}^j
\end{equation}
that, together with stochastic quantum jump operators, describes the dynamics of the system~\cite{Dalibard1992,carmichael2009open,Molmer1993}. Here, $\Gamma_0$ is the emission into free space of a single atom (we assume that the chain and impurity atoms have the same decay rate $\Gamma_0$), $k_0=2\pi/\lambda_0=\omega_0/c$, and $\bm{G}_0(\bm{r}_i,\bm{r}_j)$ is the free space Green's tensor describing the field at atom $i$ generated by atom $j$ at an energy $\omega_0$,
\begin{equation}\label{eq:greenstensor}
	\bm{G}_0(\bm{r}_i,\bm{r}_j) = \frac{1}{4\pi k_0}\left[\mathds{1}+\frac{1}{k_0^2}\nabla_i\otimes\nabla_i\right]\frac{e^{ik_0|\bm{r}_i-\bm{r}_j|}}{|\bm{r}_i-\bm{r}_j|}\,.
\end{equation}
For the terms $i=j$, we use $-3\pi G_0^{zz}(\bm{r}_i,\bm{r}_i)
=-i/2$, implicitly including the Lamb shift in the energy of the emitters. In the absence of a driving term and within the single excitation sector, as we consider here, the dynamics of the system is completely characterised by the non-hermitian Hamiltonian \eqnref{eq:fullHam}. In this case, in fact, a quantum jump prepares the system in the collective ground state which does not evolve under the action of \eqnref{eq:fullHam}.

We can separate \eqnref{eq:fullHam} into three terms including the chain, the impurity, and impurity--chain interaction parts of the Hamiltonian,
\begin{equation}\label{eq:splitHam}
	\hat{H}=\hat{H}_{\rm chain}+\hat{H}_{\rm imp}+\hat{H}_{\rm int}.
\end{equation}
The chain Hamiltonian can be written as
\begin{equation}\label{eq:Hchain_real_space}
    \Hop_{\rm chain} = \w_0 \sum_i \hat{\sigma}^i_{ee} - 3\pi \Gamma_0\sum_{i,j}G_0^{zz}(z_i,z_j)\hat{\sigma}_{eg}^i\hat{\sigma}_{ge}^j.
\end{equation}
For the case of periodic boundary conditions, and thus for infinite chains, \eqnref{eq:Hchain_real_space} can be diagonalised analytically in terms of Bloch eigenmodes $\hat{b}_k^\dagger=\frac{1}{\sqrt{N}}\sum_i e^{i k z_i}\hat{\sigma}_{\rm eg}^i$ as~\cite{porras2008collective}
\begin{equation}\label{eq:Hchain}
	\hat{H}_{\rm chain} = \sum_k \left(J_k-\frac{i}{2}\Gamma_k\right) \hat{b}_k^\dagger\hat{b}_k.
\end{equation}
In the single excitation regime, $\hat b_k$ can be taken to be bosonic annihilation operators.
Closed expressions for modes energy $J_k$ and decay rate $\Gamma_k$ are derived in Ref.~\cite{asenjo2017exponential} for the case of an infinitely long array. In the following, we are interested in the experimentally relevant case of an array with open boundary conditions. In this case, an analytical exact expression for the single excitation eigenmodes of \eqnref{eq:Hchain_real_space} is not available.  
However, when the atomic array is sufficiently long ($N\gg1$), the eigenmodes of $\hat{H}_{\rm chain}$ can still be understood as spin waves of a definite quasi-momentum $k$, where the value $k$ corresponds to the point in reciprocal space where the eigenmode wavefunction is peaked~\cite{asenjo2017exponential,porras2008collective,Mewton2007}. 
In this limit, an accurate ansatz can be provided for the single excitation eigenmodes $\bdop_k = \sum_{i} \xi_{k_\nu}(z_i)\hat{\sigma}_{eg}^i$ where~\cite{asenjo2017exponential}
\begin{equation}\label{eq:Ansatz}
	\xi_{k_\nu}(z_i) =
	\begin{cases}
	\sqrt{\frac{2}{N+1}}\,\cos(k_\nu z_i)~\text{if $\nu$ odd}\\
	\sqrt{\frac{2}{N+1}}\,\sin(k_\nu z_i)~\text{if $\nu$ even}
	\end{cases},
\end{equation}
where $k_{\nu}d=\pi\nu/(N+1)$ with $\nu=1,2,\dots,N$.
Accordingly, $J_k$ and $\Gamma_k$ in \eqnref{eq:Hchain} do not have analytically expressions, but are well approximated by the expressions for an infinite chain whenever $N\gg1$.
In the following, unless otherwise specified, we will always refer to the case of finite chains with open boundary conditions, as this is the most relevant case for the experimental realisation of atomic waveguide QED.
For $d<\lambda_0/2$, there exist single excitation eigenstates of $\hat{H}_{\rm chain}$ with 
a quasi-momentum $k>k_0$ lying outside the light cone of free space electromagnetic modes. These collective states, which are intuitively understood as excitations propagating along the array, have been shown to decay at a rate $\sim \Gamma_0/N^3$ due to scattering of the field through the ends of the array~\citep{asenjo2017exponential,Zhang2019}, and are referred to as the collective subradiant modes of the atomic array.

The second term in \eqnref{eq:splitHam}, which includes the impurity atoms and interactions among them, reads 
\begin{equation}\label{eq:Himp}
	\hat{H}_{\rm imp} = \omega_0^{\rm imp}\sum_i \hat{c}_i^\dagger\hat{c}_i + \sum_{ij} \left(g_{ij}-\frac{i}{2}\gamma_{ij}\right) \hat{c}_i^\dagger\hat{c}_j.
\end{equation}
Here we defined $\left(g_{ij}-i\gamma_{ij}/2\right)=-3\pi \Gamma_0 G_0^{zz}(\bm{r}_i,\bm{r}_j)$, and replaced the Pauli matrix of the impurity atoms with bosonic operators ($\hat{\sigma}^i_{\rm ge}\to\hat{c}_i$) as we restrict to the one-excitation dynamics~\cite{porras2008collective}. According to the definition above for the Green's tensor, when $i=j$ we have $\left(g_{ii}-i\gamma_{ii}/2\right)=-i\Gamma_0/2$.

The last term in \eqnref{eq:splitHam} contains the interactions between chain atoms and impurity atoms. Using the definitions above, the interaction Hamiltonian between the impurity atoms and the eigenmodes of the chain reads 
\begin{equation}\label{eq:Hint}
	\hat{H}_{\rm int} = \sum_{k,i}\,\tilde{g}_k^{\,i}\left(\hat{c}_i^\dagger\hat{b}_k+h.c.\right).
\end{equation}
Expanding \eqnref{eq:greenstensor} in spherical coordinates, the coupling of a single impurity atom at position $\bm{r}_i$ to a chain mode with quasi-momentum $k$ reads [see Appendix~\ref{ap:coupling}]
\begin{equation}
\tilde{g}_k^{\,i} \,= \left(g^i_k-\frac{i}{2}\gamma^i_k\right) =\, \xi_{k}(z_i)\left(|g_k^i|-\frac{i}{2}|\gamma_k^i|\right)
\end{equation}
where $\xi_{k}(z_i)$ is given in \eqnref{eq:Ansatz} and
\begin{align}\label{eq:singleatom}
	\begin{split}
	\frac{|g_k^i|-\frac{i}{2}|\gamma_k^i|}{\Gamma_0} = -\frac{3\,i}{8d}&\\
	\times\sum_{m\in\mathbb{Z}}\left[1-\kappa_m(k)^2\right]&H_0^{(1)}\left(k_0 h\sqrt{1-\kappa_m(k)^2}\right).
	\end{split}
\end{align}
Here, $\kappa_m(k)=\left(k/k_0+m\lambda_0/d\right)$, and $H_0^{(1)}$ is the Hankel function of the first kind and zeroth order. Note that for $k>k_0$, \eqnref{eq:singleatom} is purely real, which means that the coupling of an impurity to those Bloch modes is coherent. 
The coupling \eqnref{eq:singleatom} between a single impurity and a mode $k$ in the array is plotted in Fig.~\ref{fig1}(b).
For the case of an infinite array, the coupling can be obtained from \eqnref{eq:singleatom} after the substitution $\xi_{k}(z_i)\rightarrow e^{-i k z_i }/\sqrt{N}$.

Since achieving small $d$ is increasingly challenging experimentally, we consider here the case of $d=\lambda_0/4$. 
For this parameter choice, the free-space decay of the most superradiant state, $\Gamma_{k=0}$, is comparable to the width of the band $J_k$ in \eqnref{eq:Hchain}. The dynamics of an emitter with energy lying within the band of the array is thus dominated by the dissipative resonant interaction with the array's superradiant modes. We overcome this problem using as impurity, instead of single atomic emitters, the collective excitation of two neighbouring atoms, which we name atomic dimer.

\subsection{Atomic dimers}\label{sec:dimer}

We consider a dimer to be formed by two neighbouring impurity atoms at positions $\rr_i=(h,0,z_i)^T$ and $\rr_{i+1}=(h,0,z_{i+1})^T$.
For convenience, we label the two atoms forming a dimer ``$a$'' and ``$b$'' with positions $\rr_i^a=\rr_i$ and $\rr_i^b=\rr_{i+1}$, respectively.
We represent the collective single excitation of an atomic dimer by the bosonic operator
\begin{equation}\label{eq:dimerstates}
	\hat{a}^\dagger_{i\lambda}=\frac{1}{\sqrt{2}}\left(\hat{c}_i^{a\,\dagger}+\lambda\,\hat{c}_i^{b\,\dagger}\right)\,,
\end{equation}
and its Hermitian conjugate, where $\hat{c}_i^{a\,\dag}=\hat{c}_i^\dagger$ and $\hat{c}_i^{b\,\dag}=\hat{c}_{i+1}^\dagger$. For $\lambda=-1$ ($\lambda=1$), \eqnref{eq:dimerstates} creates an anti-symmetric (symmetric) excitation which predominantly couples to the sub(super)-radiant array modes exploiting their short(long)-wavelength nature. The coupling to the chain modes of the states of a dimer centered at $\rho_i=\hat{z}\cdot(\bm{r}_i^a+\bm{r}_i^b)/2$, with atoms aligned to the array atoms as depicted in Fig.~\ref{fig1}(a), writes [see Appendix~\ref{ap:coupling}]
\begin{align}\label{eq:coupdimer}
	\begin{split}
	\frac{|g_k^{i\lambda}|-\frac{i}{2}|\gamma_k^{i\lambda}|}{\Gamma_0} = -\frac{3\,i}{4\sqrt{2}\,d}&
	\left.\begin{cases}
	\sin\left(\frac{k d}{2}\right)~\text{if}~\lambda=-1\\
	\cos\left(\frac{k d}{2}\right)~\text{if}~\lambda=1
	\end{cases}\hspace{-9pt}\right\}\\
	\times\sum_{m\in\mathbb{Z}}\left[1-\kappa_m(k)^2\right]H_0^{(1)}&\left(k_0 h\sqrt{1-\kappa_m(k)^2}\right),
	\end{split}
\end{align}
and $\left(g_k^{i\lambda}-\frac{i}{2}\gamma_k^{i\lambda}\right)=\xi_k^\lambda(\rho_i)\left(|g_k^{i\lambda}|-\frac{i}{2}|\gamma_k^{i\lambda}|\right)$. This expression is similar to \eqnref{eq:singleatom} with an additional factor of $\sqrt{2}$ and a quarter-sine-wave envelope. The anti-symmetric configuration ($\lambda=-1$) eliminates the undesired dissipative coupling at small $k$ due to a destructive interference of the interaction of each atom with the chain modes, as can be seen in Fig.~\ref{fig1}(b). In the case of the dimer, coupling to half of the modes becomes zero at the center of the chain due to the particular symmetry of the collective dimer states, as indicated in \eqnref{eq:Ansatzsing} and \eqnref{eq:Ansatztrip}. Unless otherwise specified, we set $h=d$.

The dimer states in \eqnref{eq:dimerstates} are the eigenstates of \eqnref{eq:fullHam} with two atoms. Their eigenvalues are ${E_\lambda-\frac{i}{2}\Gamma_{0\lambda}}=(\omega_0^{\text{imp}}+\lambda g_{ab})-\frac{i}{2}(\Gamma_0+\lambda\gamma_{ab})$, with
\begin{equation}\label{eq:twoatoms}
\begin{split}
\frac{g_{ab}}{\Gamma_0} &= -\frac{3}{2 k_0^3r_{ab}^3}\left[\cos(k_0r_{ab})+k_0r_{ab}\sin(k_0r_{ab})\right]\\
\frac{\gamma_{ab}}{\Gamma_0} &= -\frac{3}{ k_0^3r_{ab}^3}\left[k_0r_{ab}\cos(k_0r_{ab}) -\sin(k_0r_{ab})\right]\,,
\end{split}
\end{equation}
for a separation $r_{ab}$ between the two atoms. For $r_{ab}<\lambda_0/2$, as is the case for subwavelength arrays in one dimension, the anti-symmetric dimer state has a reduced linewidth as compared to the single atom. In particular, for $d=\lambda_0/4$, the decay rate $\Gamma_{0-}\simeq\Gamma_0/4$. We have considered alternative setups, such as a 2x2 quadrupole in an anti-symmetric configuration, but the reduction of free-space decay is hindered by the reduction of its effective coupling to the chain modes [see Appendix~\ref{ap:2by2}]. We have also investigated the case in which the dimer atoms are separated by a smaller distance, for which we predict only a small improvement in performance [see Appendix~\ref{ap:smallrho}].

When the emitter's resonance energy is resonant with the subradiant part of the band, or it lies outside of the band, it is possible to adiabatically eliminate both the chain's superradiant part as well as the symmetric state of the dimer. The dynamics of the system can thus be modelled as an effective two-level system -- formed by the dimer's ground and anti-symmetric states -- coupled coherently to a set of subradiant modes [see Appendix~\ref{ap:model1}]. In the rest of the text we often refer to the resulting effective two-level system simply as dimer. Note that the expressions \eqnref{eq:coupdimer} and \eqnref{eq:twoatoms} are derived here for the case of a single dimer coupled to the array. The same derivation applies to the case of multiple dimers if these are sufficiently distant, such that the interaction in \eqnref{eq:Himp} between atoms belonging to different dimers is negligible. 

Despite the reduced effective two-level system decay rate $\Gamma_{0-}$, the dimer's linewidth is comparable to the bandwidth (BW) of the array for $d=\lambda_0/4$, leading to strong non-Markovian effects. To enter the Markovian regime, we can further reduce the dimer's linewidth using a Raman scheme as depicted in Fig.~\ref{fig1}(c), and as we now describe.

\begin{figure*}[t]
	\includegraphics[width=\linewidth]{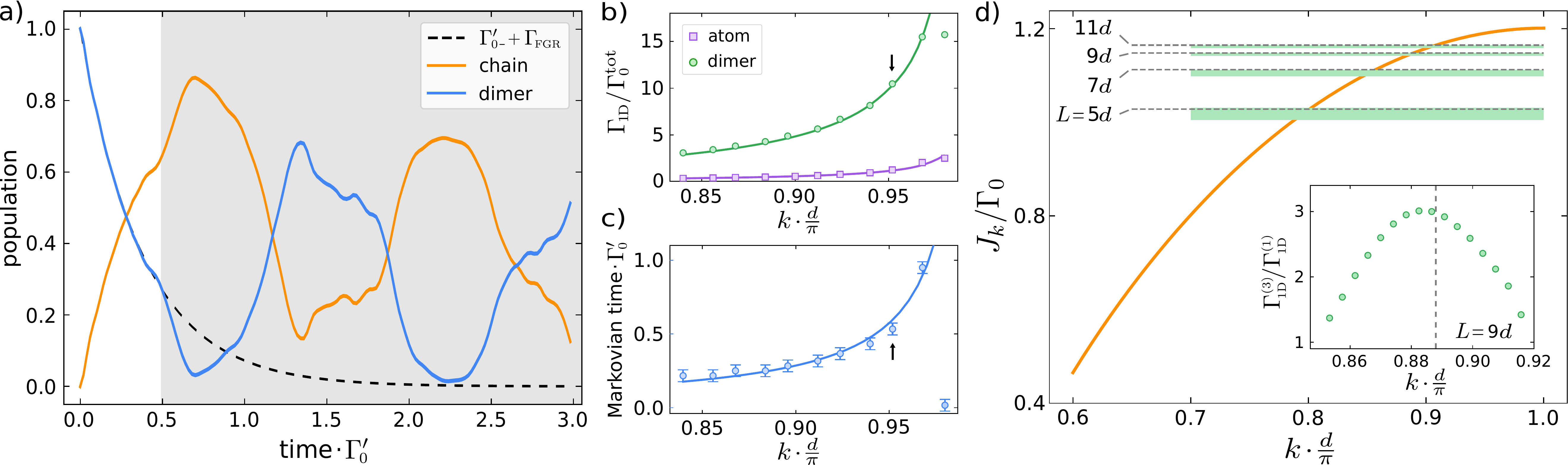}
	\caption{\label{fig2}Emitters in the band. (a) Population dynamics for a dimer anti-symmetric state resonant with $k\cdot d/\pi=0.956$. The dimer is aligned with the center of the chain. The dimer line represents the anti-symmetric state of the dimer, and the chain integrates the population of all chain modes. The population of all other states is negligible. The loss of total population is due to the dimer's free space decay. The unshaded region corresponds to the Markovian regime, in which the population is predicted by Fermi's golden rule (FGR). The shaded region corresponds to the non-Markovian regime. (b) Purcell enhancement extracted from the population dynamics. $\Gamma_{\rm 1D}$ is obtained by fitting an exponential decay to the corresponding region of the dimer's population evolution, and $\Gamma_0^{\text{tot}}$, which includes all sources of decay, by fitting the evolution of the total population in the one-excitation subspace. The solid lines show the prediction using FGR. Note the difference in performance between using a dimer or an atom. (b) and (c) The black arrow indicates the quasi-momentum of the chain mode with which the dimer in (a) is at resonance. The deviation of the points at higher $k$ from the predictions is due to the non-Markovian dynamics close to the band edge. (c) Time until the dynamics enter the shaded region in (a), measured from the numerical evolution as the time at which the population at the dimer differs from the Markovian prediction by a 6\%. The solid line shows the prediction using the length of the chain and the group velocity at $k$ to calculate the time at which the excitation in the corresponding guided mode reaches the dimer after being reflected at the ends of the chain. (d) Dispersion relation of the chain. The shaded regions indicate the energies for which the emission rate of a symmetric state of $n=3$ dimers into the chain shows $n\,\Gamma_{\rm 1D}$ superradiance for dimer--dimer separations of $L=$ 5, 7, 9, and 11$d$. Since the possible spacing between the impurities in our setup is a multiple of the array spacing, $d$, only a discrete set of energies give rise to such superradiance. The dashed lines indicate the predicted energy at which superradiance is observed, which corresponds to $k=(L-d)/L\cdot\pi/d$. The inset displays one of the plots used to obtain the green shaded regions. For $L=9$, it shows the ratio between the decay rate into the chain modes extracted from the numerical evolution with $n=3$ and the expected decay rate of a single dimer into the chain modes using FGR. Parameters for all the plots are $N=500$, $d=\lambda_0/4$, $\Delta=8~\Gamma_0$, and $\Omega=0.2~\Gamma_0$.
	}
\end{figure*}

\subsection{Raman Transition}\label{sec:Raman_transition}

In the Raman scheme, the dimer atoms are initialized in additional metastable levels $\ket{g_1^{a,b}}$, and driven into the excited state of the dipole transition included in \eqnref{eq:fullHam}, $\ket{e^i}=\hat{\sigma}_{eg}^i\ket{g_2^i}$, by a laser with Rabi frequency $\Omega$, as illustrated in Fig.~\ref{fig1}(c). The driving frequency, $\omega_R$, is detuned with respect to the energy difference between the two states by $\Delta$. The dimer atoms are excited without affecting the chain atoms by placing the chain on a node of the laser field. To describe the dynamics of the system including the Raman transition on the dimer atoms, we use the Hamiltonian in \eqnref{eq:splitHam} and include the energy of the levels $\ket{g_1^{a,b}}$, and the interaction terms due to the driving laser, $\frac{\Omega}{2}\left(\ket{g_1^\alpha}\bra{e^\alpha} e^{i\omega_R t} + h.c.\right)$. To remove the time dependence of the new interaction term, we move to a frame rotating with $\omega_R$, yielding
\begin{equation}\label{eq:Ramanevo}
\begin{split}
    \hat{H}_{\rm R} = \hat{H} + \sum_\alpha\left(\omega_0^{\rm imp}-\Delta\right)\ket{g_1^\alpha}\bra{g_1^\alpha} \\
    + \frac{\Omega}{2}\sum_\alpha\left(\ket{g_1^\alpha}\bra{e^\alpha} + h.c.\right).
\end{split}
\end{equation}
We use the evolution under the Hamiltonian in \eqnref{eq:Ramanevo}, with $\hat{H}$ being the real-space expression of \eqnref{eq:splitHam}, for all numerical results presented in this paper, which we use to benchmark the predictions obtained with analytical derivations and additional approximations.

For $\Delta,\Gamma_0\gg\Omega,g_{ab},\gamma_{ab},g_k^i,\gamma_k^i$, the states $\ket{e^{a,b}}$ remain weakly populated at all times, which we eliminate to second order in perturbation theory [see Appendix~\ref{ap:raman}]. The resulting dynamics can be approximated by an effective Hamiltonian that includes effective two-level impurity atoms $\ket{g_1^\alpha}$--$\ket{g_2^\alpha}$. For this, we define new creation operators for the dimer states, $\hat{a}^{\prime\,\dagger}_{i\lambda}=\left(\hat{c}_i^{a\prime\,\dagger}+\lambda\,\hat{c}_i^{b\prime\,\dagger}\right)/\sqrt{2}$, with $\hat{c}_i^{\alpha\prime\,\dagger}\ket{g_2^{i\alpha}} = \ket{g_1^{i\alpha}}$. We can further simplify the resulting Hamiltonian by eliminating the dimer symmetric state and the superradiant modes of the array. Using $\Delta\gg\Gamma_0$, the effective coupling between chain and dimer becomes coherent. Although we derive the effective Hamiltonian considering one dimer, the extension to multiple dimers in the case in which their interaction through free space is negligible is straightforward. Hence, for $\Delta\gg\Gamma_0\gg\Omega,g_{ab},\gamma_{ab},g_k^i,\gamma_k^i$ and distant dimers, we observe the dynamics of \eqnref{eq:Ramanevo} to be well approximated by the effective Hamiltonian
\begin{align}\label{eq:Raman}
	\begin{split}
	\hat{H}_{\rm eff} = \sum_i\left({\omega_0^{\rm imp}}^\prime-\frac{i}{2}\Gamma_{0-}^\prime\right){\hat{a}_{i-}}^{\prime\,\dagger} \hat{a}_{i-}^\prime \\
	+ \sum_{i,k>k_0} g_k^\prime\left(\xi_k^-(\rho_i)\,{\hat{a}_{i-}}^{\prime\,\dagger} \hat{b}_k + h.c.\right)\\
	+ \sum_{k>k_0} \left(J_k-\frac{i}{2}\Gamma_k\right) \hat{b}_k^\dagger \hat{b}_k \;,
	\end{split}
\end{align}
with ${\omega_0^{\rm imp}}^\prime = \omega_0^{\rm imp}-\Delta-\Omega^2/(4\Delta)$, $\Gamma_{0-}^\prime=\Omega^2/(4\Delta^2)\,\Gamma_{0-}$ and $g_k^\prime=\Omega/(2\Delta)|g_k^-|$. Note that ${g_k^\prime}^2/\Gamma_{0-}^\prime$ is independent of $\Omega$ and $\Delta$ as long as $\Delta\gg\Gamma_0\gg\Omega,g_{ab},\gamma_{ab},g_k^i,\gamma_k^i$ is fulfilled. Obtaining smaller $\Gamma_{0-}^\prime$, however, makes the Markovian regime accessible, since $\Gamma_{0-}^\prime/{\rm BW}\sim(\Omega/\Delta)^2$, while $g_k^\prime$ can be modified with other system parameters, such as tuning the resonance energy of the emitter with respect to the band, as we discuss in section \ref{sec:inband}.

In conventional waveguide QED, different physics arises when the emitter energy is inside the band of guided modes compared to when it lies in the band-gap. In the following, we consider these two regimes separately in the context of atomic dimers coupled to an atomic waveguide. For simplicity, except when the finiteness of the chain is explicitly relevant, we use the infinite array form of $\xi_k(\rho_i)$.

\section{Emitters in the band}\label{sec:inband}
\subsection{Markovian regime}

When the resonance energy of a dimer lies within the sub-radiant region of the chain's band, there is a coherent transfer of population between the dimer and the guided modes of the chain. In the Markovian regime [see the white region in Fig.~\ref{fig2}(a)], the transfer of population can be modeled as a plane wave emitted into the resonant chain mode $k$,
\begin{equation}
	\hat{H}_{\rm in}=-\frac{i}{2}\Gamma_{\rm 1D}\sum_{i,j}e^{ik(\rho_i-\rho_j)}\hat{a}_{i-}^{\prime\,\dagger}\hat{a}^\prime_{j-}-\frac{i}{2}\Gamma_{0-}^{\prime}\sum_i\hat{a}_{i-}^{\prime\,\dagger}\hat{a}^\prime_{i-},
\end{equation}
where $\Gamma_{\rm 1D}$ is the effective decay rate of the dimer excitation into the chain. A large Purcell factor $P=\Gamma_{\rm 1D}/\Gamma_{0-}^{\prime}$ corresponds to the desired regime of predominant decay of the emitters into the chain modes. The decay rate $\Gamma_{\rm 1D}$ obtained from numerical simulations agrees with the prediction using Fermi's golden rule (FGR),
\begin{equation}\label{eq:FGR}
	\Gamma_{\rm FGR} = 2Nd\frac{{g_k^\prime}^2}{\partial_kJ_k}\,.
\end{equation}
In Fig.~\ref{fig2}(b), we plot $P$ as a function of $k$ as extracted from the numerical evolution, and compare it to the prediction using FGR. We compare the case of a dimer and of a single atom, showing that the former allows for a substantial improvement in $P$ with respect to the latter. The coupling $g_k^\prime$ increases with $k$, while $\partial_kJ_k$ decreases and becomes zero at the band edge. Hence, larger $k$ are favorable and lead to a divergence in the $\Gamma_{\rm 1D}$ predicted by \eqnref{eq:FGR}. The Markovian assumption, however, breaks down at large $\Gamma_{\rm 1D}$ before reaching such divergence, as observed from the deviation between the model and the numerical results in Fig.~\ref{fig2}(b,c). The value of $P$ at a certain $k$ depends on the parameters $d$ and $h$, but not on $\Omega$ or $\Delta$. The value of $k$ at which the Markovian approximation breaks down, however, depends on $d$, $h$, and also on $\Omega/\Delta$: smaller $\Omega/\Delta$ reduces both the coupling strength and the linewidth of the dimer. Hence, the evolution stays Markovian at higher $k$, for which $P$ is larger. This improvement in $P$ is achieved at the cost of slower dynamics.

The presence of a 1D bath allows for different dimers along the chain to have a finite probability to interact with the photon emitted by the originally excited dimer. Such interaction leads to a constructive interference if $n$ dimers are prepared in a symmetric state $\hat{a}_-^{\rm sym\, \dagger}=\frac{1}{\sqrt{n}}\sum_{i=1}^n{\hat{a}_{i-}}^{\prime\,\dagger}$ and placed in an atomic mirror configuration: $|\rho_i-\rho_j|k=2\pi q$, with $q\in\mathbb{Z}$. In this state, the decay rate is enhanced by $\Gamma_{\rm 1D}^{\rm sym} = n\Gamma_{\rm 1D}$, while emission into free space is unaltered. This type of superradiance is also observed in conventional waveguide QED~\cite{Chang_2012}. In Fig.~\ref{fig2}(d), we show this feature with three dimers. Specifically, we plot the band $J_k$ and indicate which energies correspond to values of $k$ that satisfy the atomic mirror condition. The shaded regions indicate the resonance energy of the dimers for which we observe a decay rate into the chain a factor of three larger than the FGR prediction, \eqnref{eq:FGR}. We extract these regions from plots like the one that we show as an inset, in which we obtain $\Gamma_{\rm 1D}$ as a function of $k$. This collective emission can be exploited to achieve larger $\Gamma_{\rm 1D}/\Gamma_{0-}^\prime$.

\subsection{Non-Markovian regime}

The breakdown of the Markovian regime due to the diverging density of states at the edge of the band discussed before is also observable in waveguide QED setups~\cite{forn2017ultrastrong,martinez2019tunable}. In our system, we observe another type of non-Markovianity due to the finite chain length. Note that the results discussed above are independent of the number of chain atoms, $N$, except that the decay rate $\Gamma_k$ of the subradiant chain modes is slightly larger for smaller $N$. However, the length of the chain determines the time for which the dynamics stays Markovian [see the shaded region in Fig.~\ref{fig2}(a), and Fig.~\ref{fig2}(c)]. For a dimer aligned with the center of the chain, the outgoing plane waves return to the dimer after a characteristic time-scale $\tau = Nd/v_g = Nd/\partial_kJ_k$, where $v_g$ is the mode's group velocity, due to reflection at the ends of the chain. We understand these non-Markovian effects as the retarded back-action via the reflected electric field of the emitter. Mathematically, these non-Markovian effects originate from the discrete spectrum of the atomic waveguide \cite{stey1972decay}, as the dynamics of the emitter at long times is able to resolve the energy difference between two chain modes.

The evolution of the chain and dimer population in the non-Markovian regime is highly dependent on the resonance energy of the dimer, which makes the system highly tunable. We can distinguish two particular cases (i) when the dimer is resonant with an anti-symmetric chain mode and (ii) when its energy is resonant with a symmetric chain mode. The phase of the reflected wave at the dimer's position differs by a factor of $\pi$ in the two cases. In the first case, the reflected wave is in phase with the dimer and leads to an enhanced emission [note the kink in Fig.~\ref{fig2}(a)]. In the second case, the reflected wave accumulates a difference in dynamical phase. Since the coupling rate of the dimer with a symmetric mode, $\tilde{\nu}$, at the center of the chain is zero, the dimer energy effectively lies between two anti-symmetric modes. For large $N$, the energy difference with the two modes is approximately equal, $\Delta E \simeq \left|J_{k_{\tilde{\nu}\pm1}}-J_{k_{\tilde{\nu}}}\right|$, and $\tau\simeq \pi/\left|J_{k_{\tilde{\nu}\pm1}}-J_{k_{\tilde{\nu}}}\right|$, accumulating a phase difference $\varphi= \Delta E\,\tau\simeq\pi$. The reflected wave is, thus, out of phase and leads to an increase of population in the dimer [see Fig.~\ref{figNM}]. The interaction with the reflected field can also be understood as the dimer interacting with its mirror image~\citep{milonni1974retardation,dorner2002laser}, in which the non-Markovian effects are a form of retarded Dicke super or subradiance, with the emitters having (i) parallel or (ii) opposite polarization, respectively~\cite{sinha2020non}. The slow propagation of the guided modes, especially close to the band edge, enhances the retardation effects responsible of the non-Markovian behavior.

\begin{figure}[tb]
	\includegraphics[width=0.9\linewidth]{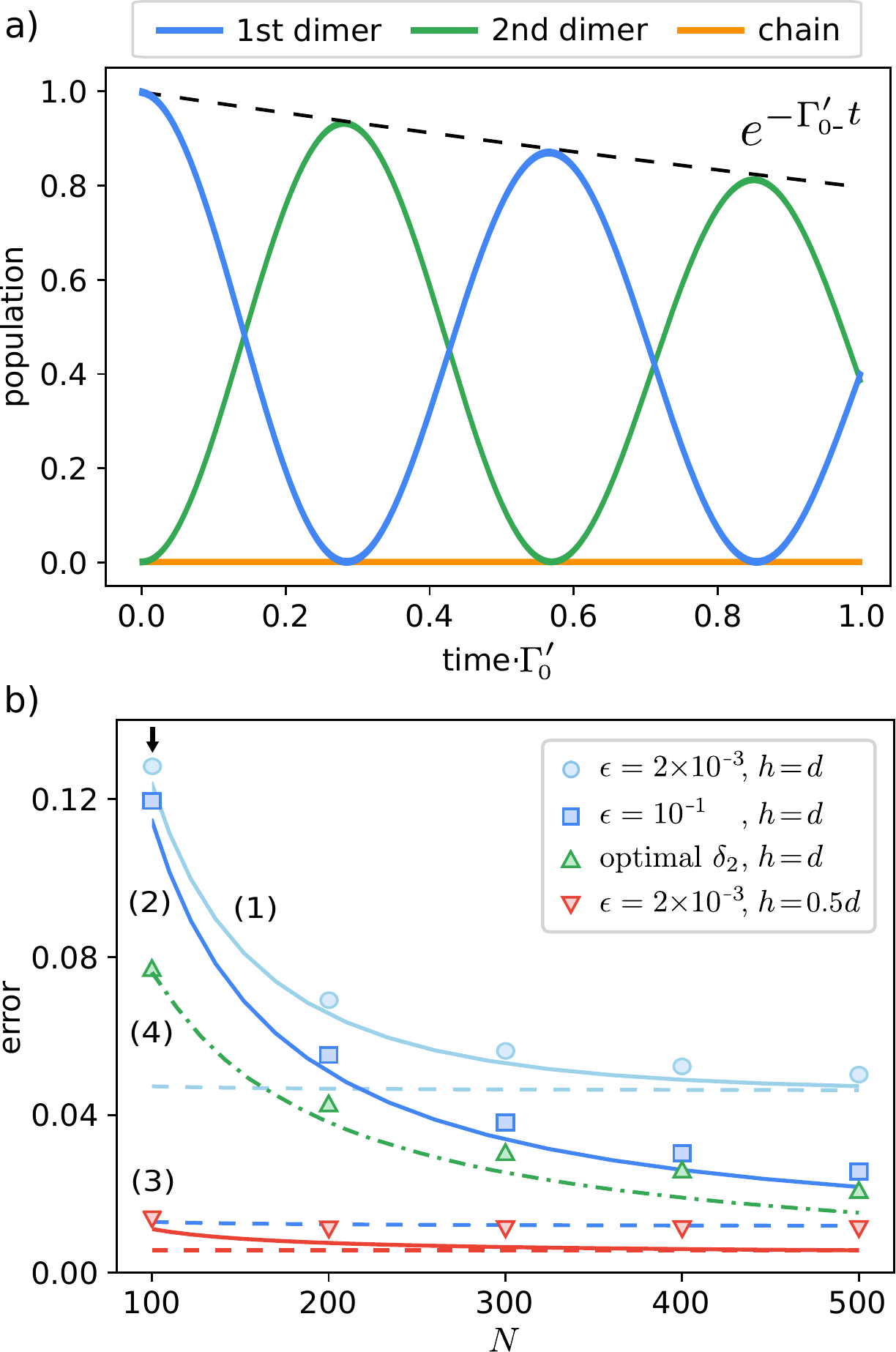}
	\caption{\label{fig3} Emitters outside the band. (a) Population dynamics for two dimers with a separation $L=14\,d$, and a detuning with the band edge $\delta=\epsilon^{-1} g_{\rm eff}^{ij}$, with $\epsilon=2\times 10^{-3}$. $N=100$. The dimer lines represent the anti-symmetric state of the dimers, and the chain integrates the population of all chain modes. The population of the rest of the states is negligible. The dashed line indicates the dimer's free space decay. (b) Error defined as the population loss in the initial dimer after one full Rabi cycle, i.e., one minus the first relative maximum of the ``1st dimer'' in (a). The black arrow indicates the point extracted from (a). The markers represent the values extracted from the numerical evolution of the full system initialized at one of the dimers' anti-symmetric state, while the lines correspond to analytical predictions for the markers with the same color.  The solid lines are computed with \eqnref{eq:LRHeff}, the dashed lines represent the continuum limit, \eqnref{eq:LRHinf}, and the dash-dotted line corresponds to \eqnref{eq:optimalerror} for an optimal $\delta_2$ given by \eqnref{eq:optimald2}.
	Parameters for all the plots are $d=\lambda_0/4$, $\Delta=200~\Gamma_0$, and $\Omega=0.03~\Gamma_0$.
	}
\end{figure}

\section{Emitters in the band-gap}\label{sec:bandgap}

If the dimer state resonance energy is located at the band-gap and the detuning with the band edge, ${\delta=\omega_0^{{\rm imp}\prime}-J_{\frac{\pi}{d}}}$, is much larger than the coupling strength to the corresponding chain modes, $g^\prime_{\frac{\pi}{d}}$, the emission into the chain is blocked. However, an atom--photon bound state with an exponentially decaying tail is formed \cite{john1990quantum}. The width of the tail scales as $1/\sqrt{\delta}$ and, for sufficiently small $\delta$, an overlap between distant atoms can be obtained \cite{douglas2015quantum,gonzalez2015subwavelength}. Adiabatically eliminating the chain modes in \eqnref{eq:Raman}, we obtain the following expression for the effective long-range coupling between distant dimers [see Appendix~\ref{ap:LR}]
\begin{equation}\label{eq:LRHeff}
	\hat{H}_{\rm eff}^{\rm LR} = \sum_{i,j,k}\frac{{g_k^\prime}^2}{{\omega_0^{\rm imp}}^\prime-J_k+\frac{i}{2}\Gamma_k}e^{ik(\rho_i-\rho_j)}\,\hat{a}_{i-}^{\prime\,\dagger}\hat{a}^\prime_{j-}.
\end{equation}
For small $\delta$, the major contributions to the sum over $k$ above are concentrated around the band edge. We thus approximate the band at the edge of the Brillouin zone as $J_{\frac{\pi}{d}(1-x)}=J_{\frac{\pi}{d}}-A_d\,x^2$, and $g_k^\prime\simeq g_{\frac{\pi}{d}}^\prime$, since $g_k$ varies slowly close to $k=\pi/d$ [see Fig.~\ref{fig1}(b)]. Likewise, the decay rate of the most subradiant modes can be approximated by $\Gamma_{\frac{\pi}{d}(1-x)}\simeq \gamma_N\,x^2$ \cite{asenjo2017exponential}, with $\gamma_N/\Gamma_0\simeq 1/N$. For subwavelength arrays, $\gamma_N/A_d < 1/N$. For compactness, we use ${\tilde{A}_d=A_d+\frac{i}{2}\gamma_N}$. With these approximations and in the continuum limit, $\Sigma_k\to\frac{Nd}{2\pi}\int dk$, we obtain a closed form for the effective coupling between two dimers mediated by the guided modes of the array,
\begin{equation}\label{eq:LRHinf}
H_{\rm eff}^{\rm LR} \simeq \sum_{i,j} g_{\rm eff}^{ij}\,e^{i\frac{\pi}{d}(\rho_i-\rho_j)}\, \hat{a}_{i-}^{\prime\,\dagger} \hat{a}^\prime_{j-} \,,
\end{equation}
with
\begin{equation}\label{eq:geff}
g_{\rm eff}^{ij} = \frac{N\pi{g_{\frac{\pi}{d}}^\prime}^2}{2\sqrt{\tilde{A}_d\,\delta}}\,e^{-\frac{\pi}{d}|\rho_i-\rho_j|/l}\,,
\end{equation}
where we identify the length scale of the interactions discussed above, $l=\sqrt{\tilde{A}_d/\delta}$. These expressions are valid as long as $\delta\gg g_{\rm eff}^{ij}$, such that elimination of the chain modes is justified. For a sufficiently large $g_{\rm eff}^{ij}$ in relation to the free-space decay, \eqnref{eq:LRHinf} predicts Rabi oscillations between the two dimers with a fidelity dictated by the ratio $g_{\rm eff}^{ij}/\Gamma_{0-}^\prime$. This ratio does not depend on the Raman transition parameters $\Delta$ and $\Omega$. However, a small $\Omega/\Delta$ allows to reduce $g_{\rm eff}^{ij}$ and, thus, one can still fulfill $\delta\gg g_{\rm eff}^{ij}\,$ with a smaller detuning $\delta$. Fixing $\epsilon= g_{\rm eff}^{ij}/\delta$, we can rewrite \eqnref{eq:geff} as
\begin{equation}\label{eq:LRcoup}
	\frac{g_{\rm eff}^{ij}}{\Gamma_{0-}^\prime} = \frac{\epsilon^{\sfrac{1}{3}}}{\Gamma_{0-}}\left(\frac{\Delta}{\Omega}\right)^{\!\!\sfrac{2}{3}}\!\left(\frac{N\pi\,\big|g_{\frac{\pi}{d}}^-\big|^2}{\sqrt{\tilde{A}_d}}\right)^{\!\!\!\sfrac{2}{3}}e^{-\frac{2\pi}{3d}|\rho_i-\rho_j|/l}\,.
\end{equation}
Note that, as expected from the continuum limit, \eqnref{eq:LRcoup} is independent of $N$, since $|g_k^-|\sim1/\sqrt{N}$, with the exception of small corrections due to a finite $\Gamma_k(N)$, that vanishes at large $N$. The dependency on $\Delta/\Omega$ indicates that the effective coupling can be made arbitrarily large at the expense of slower dynamics. We consider a system of two dimers separated by a distance $L$. Initializing the system with the excitation in one of the dimers, the chain mediates Rabi oscillations between the two otherwise non-interacting dimers, as shown in Fig.~\ref{fig3}(a) for a distance $L=14\,d$. We aim at maximising the fidelity of preparing the first dimer again in the excited state after a single Rabi cycle. The total error of the protocol, defined as the population lost to free space after one full cycle, comes from (i) the free space decay rate $\Gamma_{0-}^\prime$ of the dimers and (ii) the finite linewidth $\Gamma_k$ of the guided modes. There is also a small dephasing contribution in the exponential envelope, which we neglect in our discussion. We measure the error from the numerical evolution as one minus the first relative maximum of population at the initial dimer. Another source to the measured error is (iii) the transfer of population to the chain modes. When $\Gamma_{0-}^\prime\ll {\rm Re}[g_{\rm eff}^{ij}]$, the first contribution is approximated as $\pi\Gamma_{0-}^\prime\,/\,{\rm Re}[g_{\rm eff}^{ij}]\sim\epsilon^{-\sfrac{1}{3}}$, which is minimized for large values of $\epsilon$. The third contribution, instead, grows with $\epsilon$, as the maximum population transferred to the chain during the dynamics can be shown to be upper bound by a function proportional to $\epsilon$ [see Appendix~\ref{ap:errorchain}]. The second contribution to the error is independent of $\epsilon$, and has a value $\gamma_N/(2A_d)\sim1/N$, which sets a lower bound to the error independent of the ratio $\Omega/\Delta$.

In Fig.~\ref{fig3}(b), we plot the error for dimers interacting with chains of different lengths. We compare the numerical results with the predictions by \eqnref{eq:LRHeff} and \eqnref{eq:LRHinf}, which both include the first and second sources of error described above. We discuss four scenarios labelled (1-4) in Fig.~\ref{fig3}(b). Case (1) corresponds to the regime $\delta\gg g_k^\prime$, for which the only relevant source of error is $\Gamma_{0-}^\prime$. Case (2) corresponds to the regime $\delta\sim g_k^\prime$, in which a smaller error is obtained in expense of an exchange of population with the chain modes. Case (3) shows that placing the dimer closer to the chain as compared to the lattice spacing $d$ reduces the error thanks to the larger effective coupling rate $g_{\frac{\pi}{d}}^\prime/\Gamma_{0-}^\prime$ \cite{masson2020atomic}. The improvement is remarkable already at small $N$. Because of the discrete nature of the modes, the prediction in the continuum limit, \eqnref{eq:LRHinf}, leads to an overestimation of the resonance frequency of the modes near the band edge [see Appendix~\ref{ap:LRfiniteN}] and, hence, to the disagreement with the case of a finite chain, especially at smaller $N$. A larger effective coupling between distant dimers can, thus, be obtained by taking $\delta<0$, while staying off-resonant with the chain modes. For this, we define a new detuning between the dimer and the highest-energy array mode, $\delta_2=\omega_0^{{\rm imp}\prime}-J_{k_N}>0$. The condition $\delta_2\gg g_{\rm eff}^{ij}$ can again be made arbitrarily small by tuning $\Omega/\Delta$. For small $\delta_2$, however, $\Gamma_k$ also becomes a dominant source of error. Minimizing the error for the simplified model including only the interaction with the highest-energy chain mode, we obtain an optimal $\delta_2$ [see Appendix~\ref{ap:optimalLR}], for which we predict an error that scales as $1/N$, as in case (4) in Fig.~\ref{fig3}(b). The error deviates from the prediction at larger $N$, as the energy spacing between chain modes is reduced with $N$ and, thus, the contribution of further chain modes becomes non-negligible, for which one should go back to use \eqnref{eq:LRHeff}. Note that, although barely captured in the plots, the population in the chain modes becomes non-negligible for smaller $\delta/g_{\rm eff}^{ij}$, as in cases (2) and (4) with larger $N$.

\section{Discussion and Outlook}\label{sec:discussion}

In this work, we have proposed a setup to achieve a coherent and Markovian interaction between an emitter and the subradiant modes of an atomic chain, a system which parallels conventional waveguide QED of atoms coupled to waveguides.
Our proposal is based on two main ingredients: (i) the use of ground and anti-symmetric dimer states as an effective two-level system and (ii) the use of a Raman transition to control the linewidth of the dimer. The first method exploits the particular symmetry of the dimer state to improve the coherent coupling to the chain's guided modes by decoupling the emitter from the highly radiating modes of the chain. The second method allows to reduce the dimer's linewidth as compared to the chain's bandwidth, thus achieving the regime of Markovian dynamics. 
Accordingly, we observe similar dynamics as in conventional waveguide QED both in the in-band and band-gap regimes for the case of an atomic chain with interatomic separation of quarter wavelength. 
Along with the well-known Markovian dynamics, we also observe non-Markovian effects due to the finiteness of the chain and retardation effects introduced by the slow group velocity at the band edge.

In this work, we focused on a simple linear geometry for both the chain and the dimers, which parallels the conventional setup of atoms coupled to a linear waveguide. Several generalisations can also be considered. In particular, for the band-gap regime of emitter--chain coupling, the fidelity for coherent excitation exchange between the dimer and the chain is ultimately limited by the intrinsic decay of the chain's dark modes. 
This error can be reduced by considering emitters coupled to an atomic ring -- the atomic equivalent of a ring-resonator--, where subradiant modes are expected to have an exponential suppression of the decay rate as $\sim \Gamma_0 \exp(-N)$~\cite{asenjo2017exponential,MorenoCardoner2019}.
Furthermore, the idea of exploiting the dimer interference to reduce the coupling to the array's bright modes might be extended to 2D and 3D lattices. In higher dimensional lattices, the use of dimers (or the corresponding generalization) is expected to be particularly advantageous due to the scaling with the array's size of the superradiant modes' linewidth~\citep{porras2008collective}. 
A viable way to improve the chain-mediated coherent coupling between emitters is to consider an emitter--chain separation smaller than the interatomic separation of the chain. A reduced emitter--chain separation, which could be achieved by engineering the trapping potential using optical superlattices~\cite{Sebby-Strabley2006,Anderlini2007,Froelling2007,Trotzky2010,Lubasch2011}, is experimentally more challenging but leads to large improvements in the fidelity as shown in Fig.~\ref{fig3}.
Finally, while we only considered $d=\lambda_0/4$, our methods apply equally well to smaller interatomic separations, where the system's dynamics would benefit from the additional reduction in the decay to free-space.

This work paves the way toward observing and exploiting the rich phenomenology of waveguide QED in a clean, atom-based, setup. The additional non-Markovian effects due to the finiteness of the array are difficult to observe in standard waveguide QED and are a distinguishing feature of this platform.

\begin{acknowledgments}
D.M., C.C.R.,\ and J.I.C.\ acknowledge funding from ERC Advanced Grant QENOCOBA under the EU Horizon 2020 program (Grant Agreement No. 742102). D.C.G. acknowledges financial support from Exploring Quantum Matter.
\end{acknowledgments}

\appendix

\section{Interaction between emitters and chain modes}\label{ap:coupling}

In this appendix we outline the derivations of a closed expression for the coupling between a single atom emitter and a chain mode $k$, as in \eqnref{eq:singleatom}, and of the extension to an atomic dimer to obtain \eqnref{eq:coupdimer}. We, then, go on to discuss alternative setups.

The coupling between an impurity atom at $\bm{r}_i$ and a Bloch mode of the chain in \eqnref{eq:Hint} reads
\begin{equation}\label{eq:coupapp}
\frac{g^i_k-\frac{i}{2}\gamma_k^i}{\Gamma_0} = -\frac{3\pi}{\sqrt{N}}\sum_{j} e^{-i\bm{k}\cdot\bm{r}_j}\, G^{zz}_0(\bm{r}_i,\bm{r}_j)\,.
\end{equation}
We express \eqnref{eq:greenstensor} in cylindrical coordinates by using
\begin{equation}\label{eq:sphericalintegral}
\begin{split}
\frac{e^{ik_0|\bm{r}_i-\bm{r}_j|}}{|\bm{r}_i-\bm{r}_j|} = \frac{i}{2}\sum_{m=-\infty}^\infty\int dk\, e^{im(\phi_j-\phi_i)}e^{ik(z_j-z_i)}\\
\times J_m\left(k_\perp\rho_j\right)H_m^{(1)}\left(k_\perp\rho_i\right)\,,
\end{split}
\end{equation}
where $\rho_i>\rho_j$, and $J_m$ and $H_m^{(1)}$ are Bessel and Hankel functions of the first kind, respectively. For chain atoms sitting on the $\hat{z}$ axis, $\rho_j=0$ and $\phi_j=0$, multiplying \eqnref{eq:sphericalintegral} by $e^{-i\bm{k}\cdot\bm{r}_j}$ and summing over the array atoms yields 
\begin{eqnarray}\label{eq:periodicG}
\frac{i}{2}\sum_j \int dk^\prime&\, e^{ik^\prime(z_j-z_i)}e^{-ik z_j} &H_0^{(1)}\left(k_\perp\rho_i\right) \nonumber\\
= i\frac{\pi}{d}\sum_{m\in\mathbb{Z}}\int dk^\prime&\, \delta\left(k^\prime-k-\frac{2\pi n}{d}\right)& e^{-ik^\prime z_i}H_0^{(1)}\left(k_\perp\rho_i\right)\nonumber\\
= i\frac{\pi}{d}\sum_{m\in\mathbb{Z}}& e^{-i(k+\frac{2\pi n}{d}) z_i}\,H_0^{(1)}&\left(k_\perp\rho_i\right)
\,.
\end{eqnarray}
Plugging \eqnref{eq:periodicG} into \eqnref{eq:coupapp} yields \eqnref{eq:singleatom}.

For an even $N$, setting $z=0$ at the center of the chain,  $\bm{r}_j=dj\hat{z}$ with $j=-N+\frac{1}{2},-N+\frac{3}{2},\dots,N-\frac{1}{2}$, and for dimer's atoms in position $\bm{r}_i^\pm = \left[h\hat{\rho}+(\rho_i\pm\rho_0)\hat{z}\right]$, the coupling to the symmetric ($\lambda=1$) and anti-symmetric ($\lambda=-1$) state reads
\begin{equation}
\begin{split}
    \frac{g_k^{i\lambda}-\frac{i}{2}\gamma_k^{i\lambda}}{\Gamma_0} &= -\frac{3\pi}{\sqrt{2N}}\\
    \times\sum_j &e^{-i\bm{k}\cdot\bm{r}_j} \left[G^{zz}_0(\bm{r}_i^+,\bm{r}_j)+\lambda G^{zz}_0(\bm{r}_i^-,\bm{r}_j)\right]
\end{split}
\end{equation}

Assuming that the dimer is located far from the edges of the chain, we can extend the sum to infinite $j$ without affecting its total value. Shifting the origin to $\rho_i\hat{z}$, the sum above writes
\begin{eqnarray}
&&\sum_{j\in\mathbb{Z}+\frac{1}{2}} e^{-i\bm{k}\cdot(\bm{r}_j-\rho_i\hat{z})}\left[G^{zz}_0(\rho_0,j)+\lambda G^{zz}_0(-\rho_0,j)\right]\, \nonumber\\
&=& e^{ik\rho_i}\!\!\sum_{j\in\mathbb{Z}+\frac{1}{2}} \left[G^{zz}_0(\rho_0,j)e^{-i\bm{k}\cdot\bm{r}_j}+\lambda G^{zz}_0(-\rho_0,-j)e^{i\bm{k}\cdot\bm{r}_j}\right] \nonumber\\
&=& e^{ik\rho_i}\!\!\sum_{j\in\mathbb{Z}+\frac{1}{2}} G^{zz}_0(\rho_0,j)\left[e^{-i\bm{k}\cdot\bm{r}_j}+\lambda e^{i\bm{k}\cdot\bm{r}_j}\right]\\
&=& 2e^{ik\rho_i}\sum_{j\in\mathbb{Z}} G^{zz}_0\left(\rho_0 -d/2,j\right)\,\nonumber
\!\begin{cases}
\sin\left[k(r_j+\frac{d}{2})\right] & \!\!\!\text{if $\lambda=-1$}\\
\cos\left[k(r_j+\frac{d}{2})\right] & \!\!\!\text{if $\lambda=1$}
\end{cases},
\end{eqnarray}
where $(-)\rho_0$ and $j$ stand for $\left[\bm{r}_i^{+(-)}-\rho_i\hat{z}\right]$ and $\bm{r}_j$, respectively. For $\rho_0 = d/2$, i.e., the emitters are aligned with the chain atoms as in the text, we obtain \eqnref{eq:coupdimer}. Repeating the above treatment with the finite-chain Ansatz in \eqnref{eq:Ansatz} to replace $e^{-ikr_j}$, we derive the following definitions in the final result for $\lambda = -1$,
\begin{equation}\label{eq:Ansatzsing}
	\xi_{k_\nu}^-(\rho_i) =
	\begin{cases}
	\sqrt{\frac{2}{N+1}}\,\sin(k_\nu \rho_i)&\text{ if $\nu$ odd}\\
	\sqrt{\frac{2}{N+1}}\,\cos(k_\nu \rho_i)&\text{ if $\nu$ even}
	\end{cases},
\end{equation}
and for $\lambda = 1$,
\begin{equation}\label{eq:Ansatztrip}
	\xi_{k_\nu}^+(\rho_i)=
	\begin{cases}
	\sqrt{\frac{2}{N+1}}\,\cos(k_\nu \rho_i)&\text{ if $\nu$ odd}\\
	\sqrt{\frac{2}{N+1}}\,\sin(k_\nu \rho_i)&\text{ if $\nu$ even}
	\end{cases}.
\end{equation}

\subsection{Alternative setup: a 2x2 quadruplet}\label{ap:2by2} 

\begin{figure}[h!]
	\centering
	\includegraphics[width=0.99\linewidth]{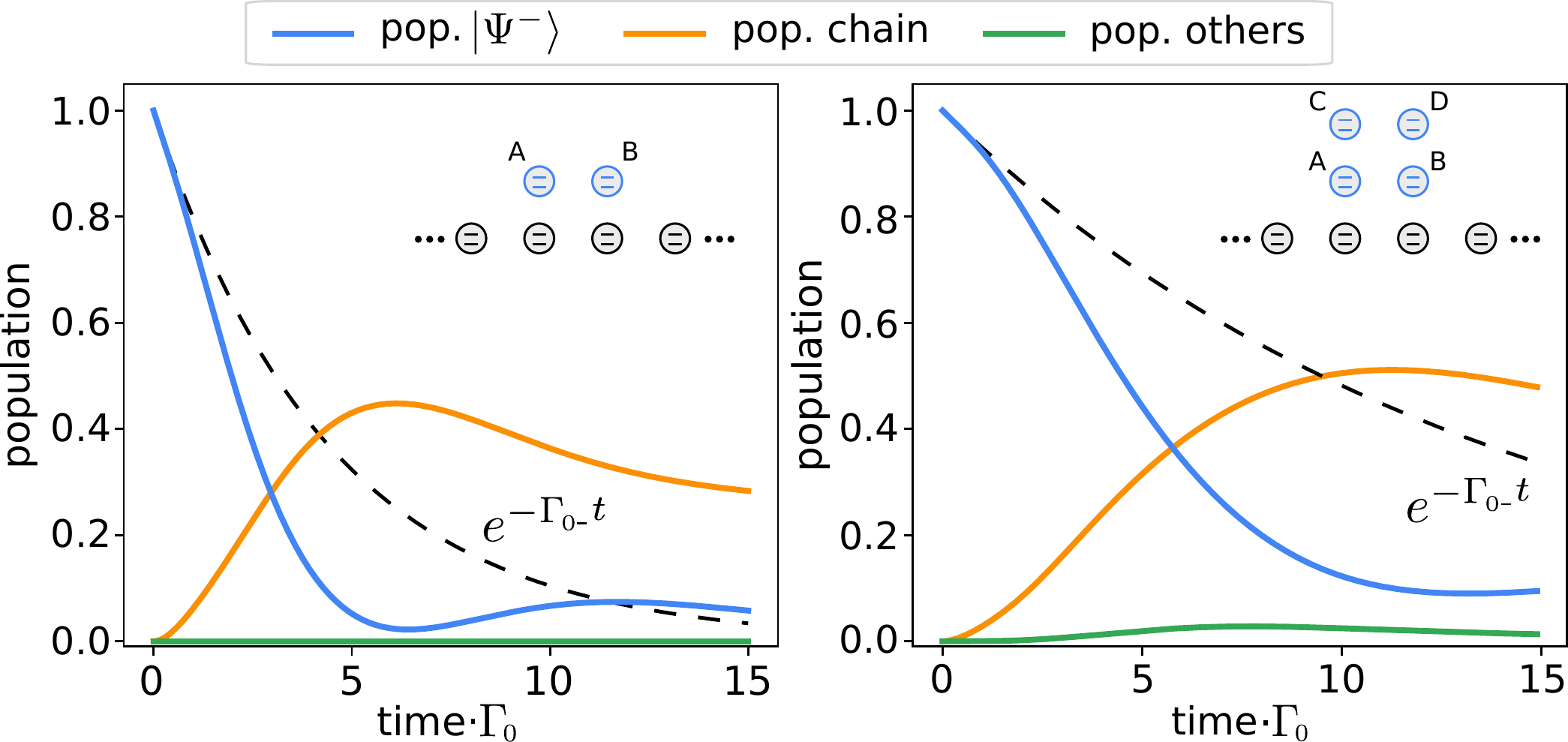}
	\caption{Dynamics of an impurity prepared in the state $\ket{\Psi^-}$ resonant with the chain subradiant modes for the case of a dimer (left panel) and a $2\times2$ plaquette (right panel). The dashed line corresponds to the free-space decay of $\ket{\Psi^-}$ at a rate $\Gamma_{0-}$. The colored lines correspond to different populations according to the legend, where \textit{chain} refers to the sum over all chain modes, and \textit{others} refers to all the eigenstates of the impurity excluding $\ket{\Psi^-}$. Parameters for this plot are $N=50$, $d=\lambda_0/4$, and the energy of the state $\ket{\Psi^-}$ is resonant with the chain mode with $k\cdot d/\pi=0.95$.}
	\label{figquad}
\end{figure}

Let us now consider a $2\times2$ plaquette of atoms and compare it to the case of a dimer considered in the main text. In general, we consider an effective two-level impurity with levels $\ket{0}$ and $\ket{\Psi^-}$. For the dimer, we use $\ket{0}=\ket{gg}$, and $\ket{\Psi^-}=(\ket{eg}-\ket{ge})/\sqrt{2}$ is the anti-symmetric state of the two atoms. For the case of a plaquette, $\ket{0}=\ket{gggg}$ is the ground state of the four plaquette atoms, and 
\begin{equation}\label{eq:psi_plaquette}
    \ket{\Psi^-}=\frac{1}{2}\left(\ket{eggg}-\ket{gegg}-\ket{ggeg}+\ket{ggge}\right)
\end{equation}
is the anti-symmetric state of the plaquette in both $x$ and $z$ directions, where the general state of the plaquette is given by $\ket{\nu_A\nu_B\nu_C\nu_D}$, with $\nu=g,e$. 
In Fig.~\ref{figquad}, we compute the evolution of the system initialised in $\ket{\Psi^-}$ in the presence of an atomic chain for both the dimer (left panel) and plaquette (right panel) configurations. We tune the atomic transition of the single impurity atoms differently for a dimer and a plaquette, such that in both cases the collective state $\ket{\Psi^-}$ has the same energy with respect to the band of subradiant modes of the array.
Fig.~\ref{figquad} shows that, while the $2\times2$ plaquette allows to attain a smaller free-space decay as compared to the dimer, it also leads to a reduction of the coupling with the chain modes and, thus, to a longer time scale for the transfer of population to the chain. After normalizing the time-axis of the two plots by the time scale of the decay rate of the respective impurity state, one can observe no qualitative difference between the evolution of the two systems.

\subsection{Alternative setup: $\rho_0<d/2$}\label{ap:smallrho}

For $\rho_0<d/2$, we can repeat the steps above to obtain
\begin{equation}\label{eq:smallrho}
\begin{split}
    \frac{|g_k^{i-}|-\frac{i}{2}|\gamma_k^{i-}|}{\Gamma_0} = \!-\frac{3\,i}{4d\sqrt{2N}}&\!\sum_{m\in\mathbb{Z}}\sin\left[k\rho_0\!+\!\left(2\frac{\rho_0}{d}-1\right)\!\pi n\right]\\
    \times \left[1-\kappa_m(k)^2\right]H_0^{(1)}&\left(k_0 h\sqrt{1-\kappa_m(k)^2}\right),
\end{split}
\end{equation}
and $\left(g_k^{i-}-\frac{i}{2}\gamma_k^{i-}\right)=\xi^-_k(\rho_i)\left(|g_k^{i-}|-\frac{i}{2}|\gamma_k^{i-}|\right)$, for the coupling rate of the anti-symmetric eigenstate of the dimer, and replacing the sine with a cosine for the symmetric state. From \eqnref{eq:twoatoms}, we know that $\gamma_{ab}$ can be further reduced by reducing $\rho_0$. \eqnref{eq:smallrho}, however, is also reduced in value for $\rho_0<0.5$. A plot of the ratio $|g^{i-}_k|^2/\Gamma_{0-}$ -- a ratio we want to optimize as we see in section \ref{sec:inband} and \ref{sec:bandgap} -- as a function of $\rho_0$ shows a slow monotonic increase when moving towards smaller $\rho_0$. The values of the ratio at $\rho_0=d/2$ and $\rho_0\to0$ for $d=\lambda_0/4$ differ by a factor of two.

\section{Effective model for dimer-chain interaction}\label{ap:model1}

In this appendix, we derive the effective Hamiltonian for the interaction betweeen the anti-symmetric state of a dimer and the subradiant modes of the chain. 
According to \eqnref{eq:splitHam}, the evolution of the dimer and of the chain within the single excitation subspace reads
\begin{equation}\label{app:eqB1}
\begin{split}
    \dot{a}_\lambda = -i\left(E_\lambda-\frac{i}{2}\Gamma_{0\lambda}\right)a_\lambda - i\sum_k \tilde{g}_k^\lambda\,b_k \\
    \dot{b}_k = -i\left(J_k-\frac{i}{2}\Gamma_k\right)b_k - i\sum_\lambda\tilde{g}_k^\lambda\,a_\lambda \,,
\end{split}
\end{equation}
where $a_\lambda$ are defined in \eqnref{eq:dimerstates}. Here, we stop using the hat notation for the operators and consider a single dimer for simplicity. The extension to multiple dimers is straightforward.

As we can see in Fig.~\ref{fig1}, interference suppresses the coupling between the anti-symmetric state of the dimer and the most dissipative modes of the chain. The same applies between the symmetric state and the subradiant modes. For a dimer anti-symmetric state energy, $E_-$, resonant with the subradiant array modes, $J_{k>k_0}$, and $d=\lambda_0/4$, the symmetric state energy, $E_+$, is resonant with the superradiant modes, $J_{k<k_0}$. In such case, $\sqrt{N}\,\tilde{g}_{k>k_0}^+\ll|J_{k>k_0}-E_+|$ and $\sqrt{N}\,\tilde{g}_{k<k_0}^-\ll|J_{k<k_0}-E_-|$. We set the zero of energy at $E_-$ and define $\Delta_+ = (E_+-E_-)$ and $\Delta_k = (J_k - E_-)$. Under these conditions and assuming that the system is initialized in the dimer anti-symmetric state, $a_-$, we adiabatically eliminate the symmetric dimer state and the superradiant modes of the chain by assuming that they acquire a negligible population, $\dot{a}_+=\dot{b}_{k<k_0}=0$, solve the corresponding equations in \eqnref{app:eqB1}, and substitute the resulting expressions for $a_+$ and $b_{k<k_0}$ in the expressions for $\dot{a}_-$ and $\dot{b}_{k>k_0}$. We obtain effective equations of motion of the form of \eqnref{app:eqB1} including only the anti-symmetric dimer state and the subradiant modes of the chain by doing the substitutions
\begin{equation}\label{eq:corrections}
    \begin{split}
        E_- &\to\; -A^-\left(1+\frac{B^*}{\Delta_+-A^+-\frac{i}{2}\Gamma_{0+}}\right) \\
        J_{k>k_0} &\to\; \Delta_{k>k_0} - \frac{\tilde{g}^{+*}_{k>k_0}}{\Delta_+ - A^+ -\frac{i}{2}\Gamma_{0+}}\sum_{k^\prime>k_0} \tilde{g}_{k^\prime}^+ \\
        \tilde{g}_{k>k_0}^- &\to\; \tilde{g}_{k>k_0}^- + \frac{B}{\Delta_+ -A^+ -\frac{i}{2}\Gamma_{0+}}\,\tilde{g}_{k>k_0}^+\;,
    \end{split}
\end{equation}
with
\begin{equation}
    A^\lambda = \sum_{k<k_0}\frac{\tilde{g}^\lambda_{k}\,\tilde{g}^{\lambda*}_{k}}{\Delta_{k}-\frac{i}{2}\Gamma_{k}}\;,\quad
    B = \sum_{k<k_0}\frac{\tilde{g}_{k}^-\,\tilde{g}_{k}^{+*}}{\Delta_{k}-\frac{i}{2}\Gamma_{k}}\;,
\end{equation}
where $\tilde{g}_k^{\lambda*} = \left(g_k^{\lambda*}-\frac{i}{2}\gamma_k^{\lambda*}\right) = \xi_k^{\lambda*}\left(|g_k^\lambda|-\frac{i}{2}|\gamma_k^\lambda|\right)$.

These corrections are negligible at the edge of the chain's band, as verified numerically by comparing the evolution under the full Hamiltonian and the effective Hamiltonian derived in this appendix. This is true because the coupling rate of the dimer anti-symmetric state with the most superradiant chain modes around $k=0$, with which the dimer is resonant due to their large linewidth, is close to zero. Since the edge of the band is our region of interest, we can hence model the system with a two-level impurity consisting of the ground and anti-symmetric state of the dimer interacting coherently with the guided ($k>k_0$) modes of the array. For lower impurity energies closer to $J_{k_0}$, the dimer becomes resonant with additional, less broad, superradiant modes with which the coupling is non-zero [see Fig.~\ref{fig2}]. In this case, \eqnref{eq:corrections} introduces finite corrections to the effective dynamics. This additional coupling to superradiant channels is further suppressed after we introduce a Raman transition in the dimer atoms due to the reduction of the dimer states' linewidth, as we see in Appendix~\ref{ap:raman}.

\section{Effective Hamiltonian with a Raman transition}\label{ap:raman}

In this appendix, we extend the effective description developed in Appendix~\ref{ap:model1} to the case of impurity atoms driven on a Raman transition (see section~\ref{sec:Raman_transition}).
We start from \eqnref{eq:Ramanevo} and separate it into a bare, $\hat{H}_0$, and an interacting, $\hat{V}$, part as
\begin{equation}
    \hat{H}_{\rm R} = \hat{H}_0 + \hat{V},
\end{equation}
with
\begin{eqnarray}
      &\hat{H}_0 = \sum_\alpha \left(\omega_0^{\rm imp}-\frac{i}{2}\Gamma_0\right)\ket{e^\alpha}\bra{e^\alpha}\nonumber\\
      &+ \sum_\alpha\left(\omega_0^{\rm imp}-\Delta\right)\ket{g_1^\alpha}\bra{g_1^\alpha} + \sum_k \left(J_k-\frac{i}{2}\Gamma_k\right) \hat{b}_k^\dagger\hat{b}_k \nonumber\\
      &+ \left(\tilde{g}_{ab}\ket{e^a}\!\bra{g_2^a}\otimes \ket{g_2^b}\!\bra{e^b}+h.c.\right),
\end{eqnarray}
and for impurity atoms at position $(h,0,z_i)^T$
\begin{equation}
    \hat{V} = \frac{\Omega}{2}\sum_\alpha\left(\ket{g_1^\alpha}\!\bra{e^\alpha} + h.c.\right) + \sum_{k,\alpha} \left(\tilde{g}_k^i\ket{e^\alpha}\!\bra{g_2^\alpha}\hat{b}_k+h.c.\right),
\end{equation}
where we use the definition of $\tilde{g}_k^*$ in Appendix~\ref{ap:model1}.

We shift the energy by $\Delta-\omega_0^{\rm imp}$, such that the excited states $\ket{e^\alpha}$ evolve fast, and define the projectors
\begin{equation}
\begin{split}
    \hat{P} =& \left(\ket{g^a_1}\!\bra{g_1^a}\otimes \ket{g_2^b}\!\bra{g_2^b} + \ket{g_2^a}\!\bra{g_2^a}\otimes\ket{g_1^b}\!\bra{g_1^b}\right)\otimes \ket{0}\!\bra{0}\\
    +&\ket{g_2^a}\!\bra{g_2^a}\otimes \ket{g_2^b}\!\bra{g_2^b}\otimes\sum_k\ket{1_k}\!\bra{1_k}\,,
\end{split}
\end{equation}
and $\hat{Q} = \mathds{1}-\hat{P}$
\begin{equation}
    = \left(\ket{e^a}\!\bra{e^a}\otimes \ket{g_2^b}\!\bra{g_2^b} + \ket{g_2^a}\!\bra{g_2^a}\otimes\ket{e^b}\!\bra{e^b}\right)\otimes \ket{0}\!\bra{0}\,,
\end{equation}
where $\hat{b}_k^\dagger\ket{0}=\ket{1_k}$. We then calculate $\hat{P}\hat{V}(\hat{Q}\hat{H}_0\hat{Q})^{-1}\hat{V}\hat{P}$ and use that, for $\Delta,\Gamma_0\gg\Omega,g_{ab},\gamma_{ab},g_k^i,\gamma_k^i$, and to second order in perturbation \cite{sternheim1972non,hung2016quantum},
\begin{equation}
    \hat{H}_{\rm eff} = \hat{P}(\hat{H}_0+\hat{V})\hat{P}-\hat{P}\hat{V}(\hat{Q}\hat{H}_0\hat{Q})^{-1}\hat{V}\hat{P}\,.
\end{equation}

\begin{figure}[h!]
	\centering
	\includegraphics[width=0.75\linewidth]{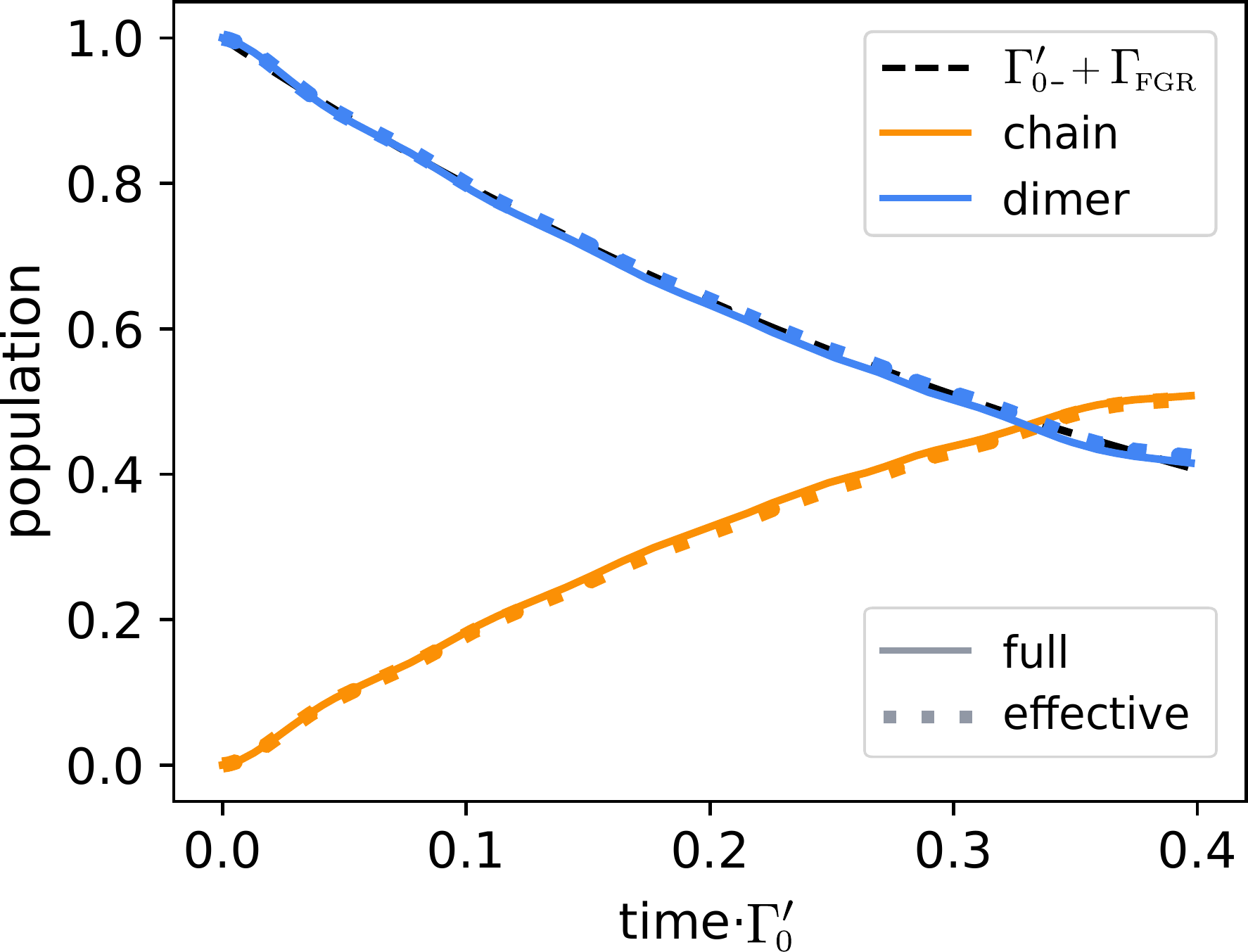}
	\caption{Population dynamics for a dimer anti-symmetric state resonant with $k\cdot d/\pi=0.945$ for $N=500$, $d=\lambda_0/4$, ${\Delta=8~\Gamma_0}$, and $\Omega=0.2~\Gamma_0$. The dimer line represents the anti-symmetric state of the dimer, and the chain integrates the population of all chain modes. The population of all other states is negligible. The loss of total population is due to the dimer's free space decay. The solid lines are computed with the full-system Hamiltonian, \eqnref{eq:Ramanevo}, and the dotted lines with the effective Hamiltonian in \eqnref{eq:Raman}.}
	\label{figRam}
\end{figure}

The resulting effective Hamiltonian in the dimer eigenstate basis and after undoing the previous energy shift reads,
\begin{widetext}
\begin{equation}\label{eq:Ramanwhole}
	\begin{split}
	\hat{H}_{\rm eff} =& \sum_{kk^\prime}\left[\left(J_k-\frac{i}{2}\Gamma_k\right)\delta_{kk^\prime} - \sum_\lambda\frac{\tilde{g}_k^{i\lambda*}\,\tilde{g}_{k^\prime}^{i\lambda}}{(\Delta+\lambda g_{ab})-\frac{i}{2}(\Gamma_0+\lambda\gamma_{ab})}\right] b_k^\dagger b_{k^\prime} + \sum_{\lambda}\left[ \omega_0^{\rm imp} - \Delta \right. \\
	& \left.- \frac{\Omega^2}{4}\frac{(\Delta+\lambda g_{ab})+\frac{i}{2}(\Gamma_0+\lambda\gamma_{ab})}{(\Delta+\lambda g_{ab})^2+\frac{1}{4}(\Gamma_0+\lambda\gamma_{ab})^2}\right]\hat{a}_{i\lambda}^{\prime\,\dagger} \hat{a}^\prime_{i\lambda} - \sum_{\lambda,k}\frac{\Omega/2}{(\Delta+\lambda g_{ab})-\frac{i}{2}(\Gamma_0+\lambda \gamma_{ab})}\left(\tilde{g}_k^{i\lambda}\,\hat{a}_j^\dagger \hat{b}_k + h.c.\right) \;.
	\end{split}
\end{equation}
\end{widetext}
In the regime $\Delta\gg\Gamma_0\gg\Omega,g_{ab},\gamma_{ab},g_k^i,\gamma_k^i$, the dimer energy and the dimer--chain coupling can be approximated as in \eqnref{eq:Raman}. For the dimer energy resonant with the subradiant region of the chain's band, we can eliminate the symmetric dimer state and the superradiant chain modes in the same way than in Appendix~\ref{ap:model1}. Note that with the Raman transition, the symmetric state is also resonant with the subradiant modes. Nevertheless, its coupling is smaller and its linewidth larger than the one of the anti-symmetric state, and we verify that it can be safely neglected when studying the dynamics of the latter. In contrast to Appendix~\ref{ap:model1}, the reduced dimer linewidth due to the Raman transition maintains the correction due to the superradiant chain modes small at energies close to $J_{k_0}$. The additional $\sim g_k^2$ term in \eqnref{eq:Ramanwhole} introduces small shifts in the energy of the chain modes. We account for them by correcting the $k$ axes in Fig.~\ref{fig2} with the momentum extracted from the chain mode being excited, which might not coincide exactly with the $k$ expected from $J_k$. In Fig.~\ref{figRam}, we compare the evolution of a dimer anti-symmetric state resonant with the band of subradiant chain modes computed both with the full Hamiltonian in \eqnref{eq:Ramanevo}, and the effective Hamiltonian in \eqnref{eq:Raman}, and verify that the approximations discussed here are reasonable.

\section{Non-Markovian dynamics due to retardation and the finiteness of the array for a dimer resonant with a symmetric array mode}\label{ap:NM}

\begin{figure}[h!]
	\centering
	\includegraphics[width=0.8\linewidth]{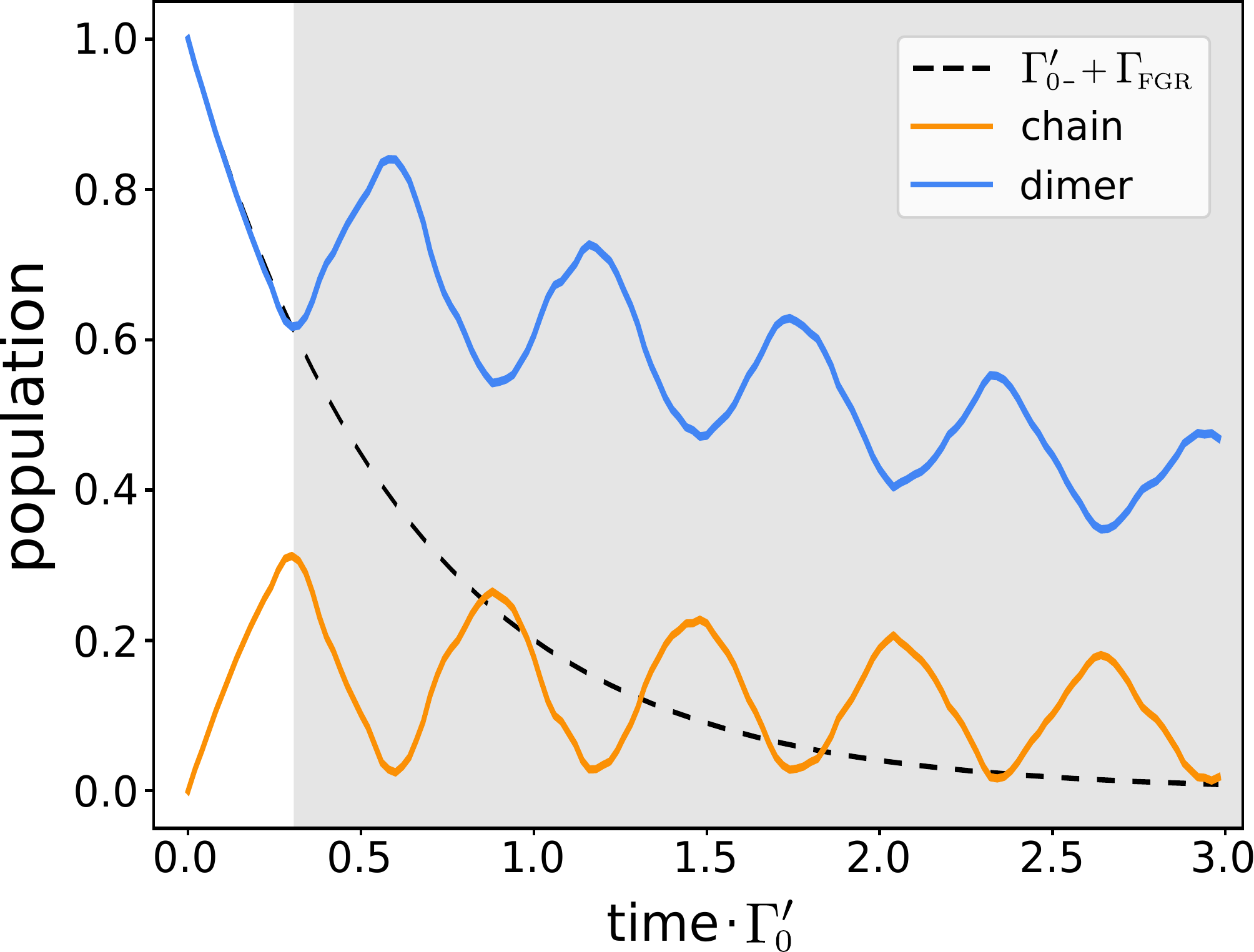}
	\caption{Population dynamics for a dimer anti-symmetric state resonant with $k\cdot d/\pi=0.922$ for $N=500$, $d=\lambda_0/4$, ${\Delta=8~\Gamma_0}$, and $\Omega=0.2~\Gamma_0$. The dimer is aligned with the center of the chain. The dimer line represents the anti-symmetric state of the dimer, and the chain line integrates the population of all chain modes. The population of all other states is negligible. The loss of total population is due to the dimer's free space decay. The white region corresponds to the Markovian regime, in which the population is predicted by Fermi's golden rule. The shaded region corresponds to the non-Markovian regime.}
	\label{figNM}
\end{figure}

In this appendix, we show the non-Markovian dynamics of a dimer relaxing into an atomic chain when the dimer is resonant with a symmetric array modes.
At long times, comparable to one over the energy difference between the discrete guided modes of the array, we observe non-Markovian effects in the dynamics of a dimer with energy inside the chain's band. We understand these effects as the retarded back-action via the electric field of the emitter, which is reflected back at the ends of the chain. For a dimer aligned with the center of the chain, and with its anti-symmetric state energy resonant with a chain mode with even symmetry, such as $k\cdot d/\pi=0.922$ in Fig.~\ref{figNM}, the coupling rate to that mode is zero. Therefore, the dimer's energy effectively lies between $k\cdot d/\pi=0.920$ and $k\cdot d/\pi=0.924$. The reflected wave is, thus, out of phase with the dimer, as discussed in the text, and leads to an increase of the dimer's population, as we see when the evolution enters the shaded area in Fig.~\ref{figNM}.

\section{Effective dimer--dimer long range interaction in the band-gap}\label{ap:LR}

In this appendix, we provide additional detail to the case of a multiple dimers interacting with an atomic chain when the dimers' energy lies within the chain's band-gap. In particular, we derive the effective model for the dimers-chain evolution in this regime, we explain the origin of the discrepancy between the case of finite and infinite long chain, and we derive the optimal error's scaling for the excitation swapping between dimers mediated by the chain.

The evolution of the operators $a^\prime_{i-}$ and $b_k$ under the Hamiltonian in \eqnref{eq:Raman} is
\begin{equation}
    \begin{split}
        \dot{a}^\prime_{i-}(t) &= -i\left({\omega_0^{\rm imp}}^\prime-\frac{i}{2}\Gamma_{0-}^\prime\right)a^\prime_{i-} -i\sum_k g_k^\prime\, e^{ik\rho_i}\, b_k \\
\dot{b}_k(t) &= -i\left(J_k-\frac{i}{2}\Gamma_k\right)b_k - i\sum_i g_k^\prime\, e^{-ik\rho_i}\, a^\prime_{i-}  \;.
    \end{split}
\end{equation}
Moving to a rotating frame, $\tilde{a}^\prime_{i-}(t) = a^\prime_{i-}(t)\,e^{i{\omega_0^{\rm imp}}^\prime t}$ and $\tilde{b}_k(t) = b_k(t)\,e^{iJ_k t}$, solving for $\tilde{b}_k(t)$, and substituting into the equation for $\dot{\tilde{a}}^\prime_{i-}(t)$, we obtain an effective equation for the dimer's population,
\begin{equation}
\begin{split}
    \dot{\tilde{a}}^\prime_{i-}(t) &= -\frac{\Gamma_{0-}^\prime}{2}\tilde{a}^\prime_{i-} - i\sum_{k,j} {g_k^\prime}^2\,e^{ik(\rho_j-\rho_i)}\\
    \times&\int_0^t e^{i\left[{\omega_0^{\rm imp}}^\prime-\left(J_k-\frac{i}{2}\Gamma_k\right)\right](t-t^\prime)} \tilde{a}^\prime_{j-}(t^\prime) \,dt^\prime \;.
\end{split}
\end{equation}
We assume that, after summing over momenta, the time integral only contributes for a small correlation time $\tau_c$ (Markov approximation). Since $g_k^\prime$ is approximately constant close to the band edge, the region that is closest in resonance with the dimers, $\tau_c$ is short. Assuming that the dimer operator evolves over time scales much longer than $\tau_c$, we approximate $\tilde{a}_{i-}^{\prime\,\dagger}(t^\prime)\simeq\tilde{a}_{i-}^{\prime\,\dagger}(t)$. For $t\gg\tau_c$ and $\omega_{\rm at}^\prime-J_{\pi/d} > 0$, the equation for the evolution of the dimer population reads
\begin{equation}
\begin{split}
    \dot{a}^\prime_{i-}(t) =& -i\left({\omega_0^{\rm imp}}^\prime -\frac{i}{2}\Gamma_{0-}^\prime\right) a^\prime_{i-}(t) \\
    -&i\sum_{k,j} \frac{{g_k^\prime}^2}{{\omega_0^{\rm imp}}^\prime-\left(J_k-\frac{i}{2}\Gamma_k\right)}\,e^{ik(\rho_j-\rho_i)}\, a^\prime_{j-}(t) \;,
\end{split}
\end{equation}
from which we can infer the effective dimer--dimer interaction in \eqnref{eq:LRHeff}.

The biggest contributions to $g_{\rm eff}^{ij}$ come from the $k$ near the band edge. To obtain a closed form of the effective coupling between dimers through the guided modes of the chain, we approximate $g_k^\prime\simeq g_{\frac{\pi}{d}}^\prime$, and do a quadratic band approximation of $J_k$ around the band edge. Expanding the analytical form of the dispersion relation for an infinite chain \cite{asenjo2017exponential} around $k=\pi/d$ and truncating to second order, we obtain $J_{\frac{\pi}{d}(1-x)}\simeq J_{\frac{\pi}{d}}- A_d\,x^2$, with
\begin{equation}
      \frac{A_d}{\Gamma_0} = \frac{3\pi^2}{2k_0^3d^3}\bigg[\log\left[2\cos\left(\frac{k_0d}{2}\right)\right] + \frac{k_0d}{2}\tan\left(\frac{k_0d}{2}\right)\bigg],
\end{equation}
for $d<\lambda_0/2$. For $d=\lambda_0/4$, $A_d/\Gamma_0\simeq 4.3$. Defining $\mu\in[1,N]$, such that the discrete $k_\mu= \frac{\pi}{d}\left(1-\frac{\mu}{N+1}\right)$, the decay rate of the most subradiant modes scales as ${\Gamma_\mu\simeq \gamma_N\mu^2/N^2}$ \cite{asenjo2017exponential}. Thus, $\Gamma_{\frac{\pi}{d}(1-x)}\simeq \gamma_N x^2$, and in the continuous band approximation,
\begin{equation}\label{eq:Lraproxs}
\begin{split}
    \sum_k\!&\frac{{g_k^\prime}^2\,e^{ik(\rho_i-\rho_j)}}{{\omega_0^{\rm imp}}^\prime-J_k+\frac{i}{2}\Gamma_k} \\
&\simeq \frac{N{g_{\frac{\pi}{d}}^\prime}^2}{2\,\delta}\!\!\int_{-1}^1\!\!dx\, \frac{e^{i\frac{\pi}{d}(\rho_i-\rho_j)(1-x)}}{1+\frac{1}{\delta}\left(A_d+\frac{i}{2}\gamma_N\right)x^2},
\end{split}
\end{equation}
where $\delta= {\omega_0^{\rm imp}}^\prime-J_{\frac{\pi}{d}}$. For $\delta\ll A_d$, the integration limits can be extended to infinity without affecting the solution. We introduce $\eta$ by substituting $(1-x)$ with $(\eta-x)$. Using the convolution theorem $(f\ast g)(\eta) = \mathcal{F}^{-1}\{\mathcal{F}(f)\cdot\mathcal{F}(g)\}$, and the results of the Fourier transform ${\mathcal{F}(e^{ik\rho\,x})(\nu) = \delta(k\rho-\nu)}$, and ${\mathcal{F}\left((1+A\,x^2)^{-1}\right)(\nu) = \frac{\pi}{\sqrt{A}}e^{-|\nu|/\sqrt{A}}}$, \eqnref{eq:Lraproxs} becomes
\begin{equation}
\begin{split}
    &\sum_k\!\frac{{g_k^\prime}^2\,e^{ik(\rho_i-\rho_j)}}{{\omega_0^{\rm imp}}^\prime-J_k+\frac{i}{2}\Gamma_k} \simeq\frac{N\pi}{2}\frac{{g_{\frac{\pi}{d}}^\prime}^2}{\delta}\sqrt{\frac{\delta}{A_d+\frac{i}{2}\gamma_N}}\\
    &\times\int_{-\infty}^\infty d\nu\, e^{i\nu\eta}\,\delta\left(\frac{\pi}{d}(\rho_i-\rho_j)-\nu\right) e^{-|\nu|\sqrt{\delta/\left(A_d+\frac{i}{2}\gamma_N\right)}}\,,
\end{split}
\end{equation}
and \eqnref{eq:LRHinf} and \eqnref{eq:geff} follow trivially. Note that the exponential envelope in real space introduces a length scale for the interactions: $l=\sqrt{A_d/\delta}$.

\subsection{Error due to populating the chain modes}\label{ap:errorchain}
In the regime in which $\delta\gg g_{\frac{\pi}{d}}^\prime$ is not satisfied, but $\epsilon=g_{\frac{\pi}{d}}^\prime/\delta$ is still small, the transition probability at time, $t$, from an initial dimer state with energy $J_{\frac{\pi}{d}}+\delta$ to the chain modes, to first order
\cite{cohen1998atom}
\begin{equation}
    \mathcal{P}(t) = 4\,g_{\frac{\pi}{d}}^{\prime\,2}\,\sum_k \frac{\sin^2\left[(J_{\frac{\pi}{d}}+\delta-J_k)\,t/2\right]}{\left(J_{\frac{\pi}{d}}+\delta-J_k\right)^2}\,.
\end{equation}
In the continuum limit, $\Sigma_k\to\frac{Nd}{2\pi}\int dk$, we extract an expression for the absolute maximum of population at the chain
\begin{equation}
    {\rm max}[\mathcal{P}(t)] \simeq \frac{5}{2}N\,\epsilon^{\sfrac{3}{2}}\,\sqrt{\frac{g_{\frac{\pi}{d}}^\prime}{A_d}}
\end{equation}
Note that this probability is an upper bound both because it is the maximum of population at the chain, and because the integral in the continuum limit overestimates the value of the sum over $k$ for a small $\delta$ compared to $J_{\frac{\pi}{d}}-J_{\frac{\pi}{d}(1-1/N)}$, as discussed in section \ref{sec:bandgap} and Appendix~\ref{ap:LRfiniteN}. Since now we compute $1/\delta^2$, the mismatch between the results from using a continuum or a discrete band scales faster than in the calculation of $g_{\rm eff}^{ij}$. Finally, the error defined in the text is not well suited to capture error due to populating the chain modes, as the time scale of the oscillations between chain and dimers is much shorter than the one of the oscillations between dimers. The error at the dimer's maximum is, therefore, most likely measured at a moment in which the population in the chain is zero, and the probability that such point of time exists close to the maximum of dimer population is high.

\subsection{Prediction mismatch due to the finite $N$}\label{ap:LRfiniteN}
\begin{figure}[h!]
	\centering
	\includegraphics[width=0.99\linewidth]{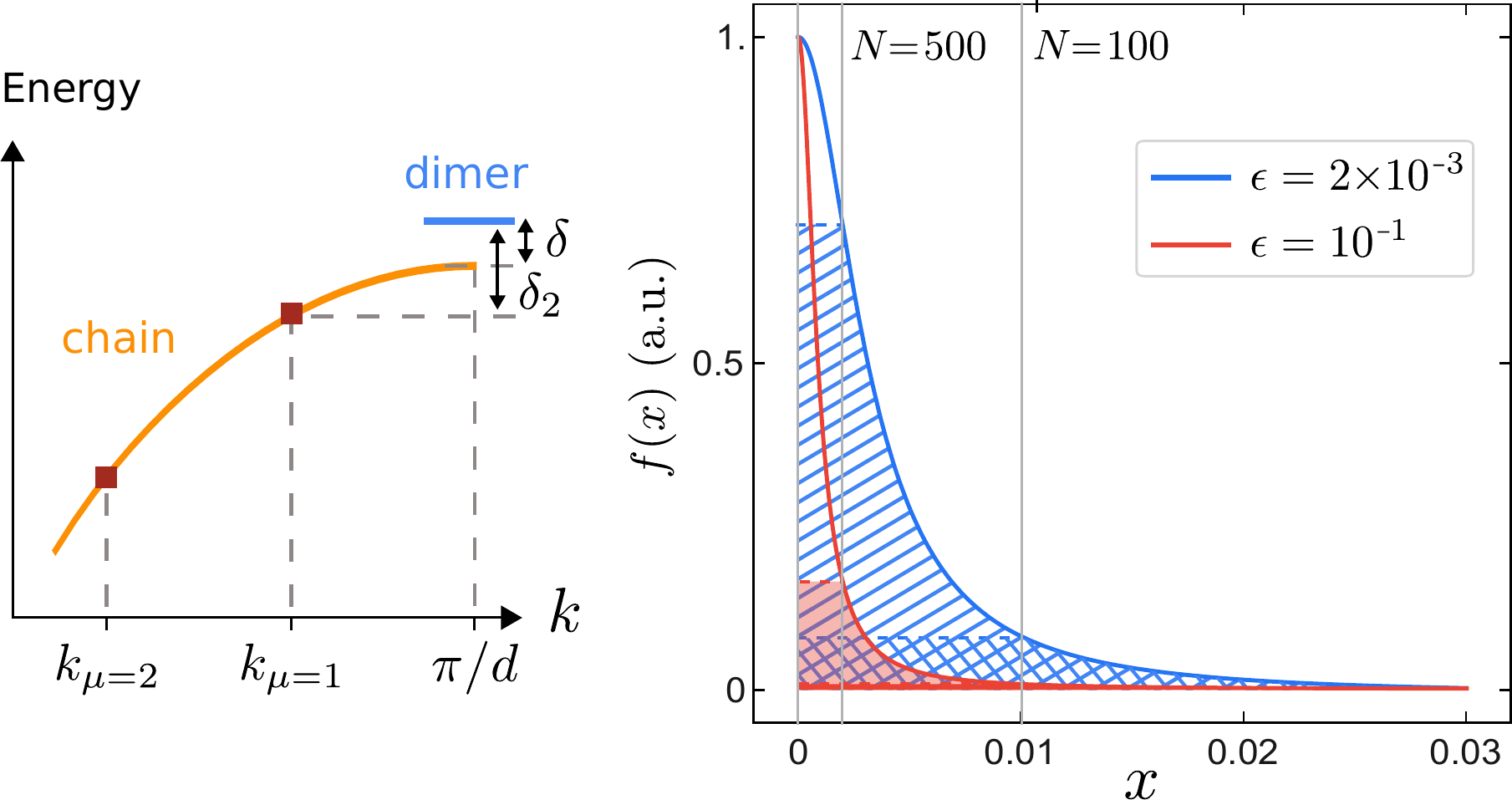}
	\caption{Left panel: Sketch of the energy of the system consisting of emitters in the chain's band-gap discussed in section \ref{sec:bandgap}. The solid line represents the chain's band in the continuum limit, while the markers point at the discrete levels of a finite chain. The dimer energy is detuned by $\delta$ with the band edge, and by $\delta_2$ with the highest-energy discrete state. Right panel: $f(x)$ [see \eqnref{eq:fx}] for the different values of $\epsilon=g_{\rm eff}^{ij}/\delta$ used in Fig.~\ref{fig3}. The three vertical lines indicate, from left to right, $x=0$, and the value of $x$ corresponding to $k_{\mu=1}$ [~$x=\mu/(N+1)$~] for $N=500$ and $N=100$, respectively. The shaded regions highlight the area under $f(x)$ obtained by integrating between $x=1$ and $x=1/(N+1)$, notice $k_{\mu=1}$, and using the discrete sum from that point to $x=0$. Parameters used are $d=\lambda_0/4$, ${\Delta=200~\Gamma_0}$, and $\Omega=0.03~\Gamma_0$.}
	\label{figMismatch}
\end{figure}

In Fig.~\ref{fig3}, we observe a difference between the effective dimer--dimer interaction as described by \eqnref{eq:LRHeff} and \eqnref{eq:LRHinf} for a dimer' energy in the band-gap, as depicted in Fig.~\ref{figMismatch} (left). This disagreement is due to the discreteness of the chain modes, for which the highest-energy mode has quasi-momentum $k_{\mu=1}$ [see Appendix~\ref{ap:LR}], with $k_{\mu=1}\simeq\frac{\pi}{d}(1-1/N)$. Since the biggest contributions to $g_{\rm eff}^{ij}$ come from the modes closest in energy to the band edge, and $J_{k_{\mu=1}}<J_{\frac{\pi}{d}}$, the larger detuning between the dimer and the mode with $k_{\mu=1}$ ($\delta_2$ in Fig.~\ref{figMismatch}) explains the smaller coupling rates predicted by \eqnref{eq:LRHeff}. In other words, the approximation in \eqnref{eq:Lraproxs} may not be appropriate below a certain $N$.

We can better understand the mismatch between the discrete and infinite chain predictions by looking at the integrand in \eqnref{eq:Lraproxs}
\begin{equation}\label{eq:fx}
    f(x) = \frac{\cos\left[\frac{\pi}{d}(\rho_i-\rho_j)x\right]}{1+\frac{1}{\delta}A_d\, x^2}\,,
\end{equation}
as we do in Fig.~\ref{figMismatch} (right). For simplicity, we focus the comparison on the most dominant term of the discrete sum, the one with $k_{\mu=1}$, for which the difference in total value between the two sides of \eqnref{eq:Lraproxs} is captured by the non-shaded areas in Fig.~\ref{figMismatch}. For instance, the ratio between the non-shaded part of the area under $f(x)$ and the full integral is rather small for $\epsilon = 2\times 10^{-3}$ and $N=500$, and the offset between the corresponding marker and dashed line in Fig.~\ref{fig3} is also small. However, the steepness of the function at small $x$ can lead to large underestimations of the area under the function. This becomes critical at smaller $N$. For $\epsilon = 2\times 10^{-3}$ and $N=100$, the non-shaded part amounts to multiple times the shaded part of the area under $f(x)$. The value of the mismatch between the discrete and continuum predictions also depends on $\delta$, as smaller $\delta$ are better able to resolve the detuning of the mode $k_{\mu=1}$ with the band edge. In other words, $f(x)$ becomes steeper, and the amount of underestimation of the integral using the discrete sum is larger, as shown in Fig.~\ref{figMismatch} for $\epsilon=10^{-1}$. This, again, can be verified by observing Fig.~\ref{fig3}.

The predictions with \eqnref{eq:LRHinf} are, thus, in general overly optimistic. However, by studying the discrete spectrum, we can optimize the predicted coupling rates, as we do in the following section.

\subsection{Optimal error for $\delta<0$}\label{ap:optimalLR}

Analyzing the system from the picture of a discrete spectrum of array modes with momenta $k_\mu$, as defined above and sketched in Fig.~\ref{figMismatch} (left), instead of a continuous band, we can improve the long-range coupling rate between dimers. In this case, the expression for the effective dimer--dimer coupling reads
\begin{equation}
	g_{\rm eff}^{ij} \simeq g_{\frac{\pi}{d}}^{\prime\,2} \sum_{\mu=1}^N \frac{\xi^-_{k_\mu}(\rho_i)\,\xi_{k_\mu}
	^{-*}(\rho_j)}{\delta+\left(A_d+\frac{i}{2}\gamma_N\right)\frac{\mu^2}{N^2}}\;.
\end{equation}
Defining a detuning $\delta_2$ from the highest energy mode of the array ($\mu=1$),
\begin{equation}\label{eq:LRdiscrete}
	g_{\rm eff}^{ij} \simeq g_{\frac{\pi}{d}}^{\prime\,2} \sum_{\mu=1}^N \frac{\xi^-_{k_\mu}(\rho_i)\,\xi_{k_\mu}^{-*}(\rho_j)}{\delta_2+\frac{i}{2}\gamma_N\frac{\mu^2}{N^2}+\frac{A_d}{N^2}\left(\mu^2-1\right)}\;.
\end{equation}
To stay off-resonant with the mode $k_{\mu=1}$, we need $\delta_2>\frac{\gamma_N}{N^2}$. For $N\gg1$, $(\delta_2-J_{k_2})= \frac{3A_d}{N^2}\gg\frac{\gamma_N}{N^2}\sim \delta_2$. Thus, we approximate \eqnref{eq:LRdiscrete} with the first term of the sum,
\begin{equation}
	g_{\rm eff}^{ij} \,\simeq\, g_{\frac{\pi}{d}}^{\prime\,2}\,\frac{\delta_2-i\frac{\gamma_N}{2N^2}}{\,\;\delta_2^2+\left(\frac{\gamma_N}{2N^2}\right)^2}\;,
\end{equation}
where for simplicity we have also used $\xi^-_{k_1}(\rho_i)\xi_{k_1}^{-*}(\rho_j)\simeq 1$, valid for a long chain and dimers located near its center.

As discussed in section \ref{sec:bandgap}, the dimer--dimer long-range coupling has three main sources of error. One is due to $\Gamma_{0-}^\prime$, given by the decay of the dimer population after one full Rabi cycle, $\pi\,\Gamma_{0-}^\prime/{\rm Re}[g_{\rm eff}^{ij}]$. A second source is due to the small but finite linewidth of the array modes that mediate the interaction, $\Gamma_k$, and write $2\pi\,{\rm Im}[g_{\rm eff}^{ij}]/{\rm Re}[g_{\rm eff}^{ij}]$. Thirdly, the finite population transfer to the array mode if $\delta_2\gg g_{\frac{\pi}{d}}^\prime$ is not satisfied. The maximum of population in the chain is $4g_{\frac{\pi}{d}}^{\prime\,2}/(\delta_2^2+4g_{\frac{\pi}{d}}^{\prime\,2})$.

The error in Fig.~\ref{fig3} is due to the free-space decay of the dimers and that of the array modes and reads
\begin{equation}
    {\rm error} \;\simeq \;\frac{\pi\Gamma_{0-}}{\big|g_{\frac{\pi}{d}}^-\big|^2}\,\delta_2\; +\; \pi\frac{\gamma_N}{N^2}\,\frac{1}{\delta_2}\,.
\end{equation}
By minimizing the error, we obtain an optimal value of $\delta_2$
\begin{equation}\label{eq:optimald2}
    \delta_2^{\rm opt} = \frac{\big|g_{\frac{\pi}{d}}^-\big|}{\sqrt{N^2\,\Gamma_{0-}/\gamma_N}}\,,
\end{equation}
for which we predict an error
\begin{equation}\label{eq:optimalerror}
    {\rm error}_{\rm opt}\simeq \frac{2\pi}{\big|g_{\frac{\pi}{d}}^-\big|}\frac{\sqrt{\Gamma_{0-}}}{ N^{\sfrac{3}{2}}}\,.
\end{equation}
Since $\big|g_{\frac{\pi}{d}}^-\big|$ scales as $1/\sqrt{N}$, the optimal error scales as $1/N$.

\bibliography{bibliography}

\end{document}